\documentclass[bibyear]{aa}
\usepackage{graphicx}

\def\MF{\mbox{CH$_3$OCHO}}
\def\MCN{\mbox{CH$_3$CN}}
\def\MCNI{$^{13}$CH$_3$CN}
\def\MCNII{CH${_3}^{13}$CN}

\def\HII{H{\sc ii} }

\def\UC{UC~H{\sc ii}}

\def\kms{\mbox{km~s$^{-1}$}}

\def\Vlsr{$V_{\rm LSR}$}

\def\mjy{~mJy~beam$^{-1}$}

\def\g31{G31}

\begin{document}
\title{Accelerating infall and rotational spin-up in the hot molecular core G31.41+0.31} 
\author{M.\ T.\ Beltr\'an\inst{1}, R.\ Cesaroni\inst{1}, V.\ M.\ Rivilla\inst{1}, \'A.\ S\'anchez-Monge\inst{2}, L.\ Moscadelli\inst{1}, A.\ Ahmadi\inst{3}, V.\ Allen\inst{4, 5}, H.\ Beuther\inst{3}, S.\ Etoka\inst{6}, D.\ Galli\inst{1}, R.\ Galv\'an-Madrid\inst{7}, C.\ Goddi\inst{8, 9}, K.\ G.\ Johnston\inst{10}, P.\ D.\ Klaassen\inst{11}, A.\ K\"olligan\inst{12},  R.\ Kuiper\inst{12}, M.\ S.\ N.\ Kumar\inst{13, 14}, L.\ T.\ Maud\inst{15, 9},  J.\ C.\ Mottram\inst{3},  T.\ Peters\inst{16}, P.\ Schilke\inst{2},  L.\ Testi\inst{1, 17}, F.\ van der Tak\inst{4, 5}, C.\ M.\ Walmsley\thanks{Malcolm Walmsley could not see the completion of this article, as he passed away a few months ago. However, his contribution to our project was crucial and will never be forgotten.}
}
\institute{
INAF-Osservatorio Astrofisico di Arcetri, Largo E.\ Fermi 5,
I-50125 Firenze, Italy
\and
I.\ Physikalisches Institut, Universit\"at zu K\"oln, 
Z\"ulpicher Str.\ 77, D-50937 K\"oln, Germany
\and
Max Planck Institute for Astronomy, K\"onigstuhl 17, 69117 Heidelberg, Germany
\and
Kapteyn Astronomical Institute, University of Groningen, 9700 AV, Groningen, The Netherlands
\and
SRON Netherlands Institute for Space Research, Landleven 12, 9747 AD, Groningen, The Netherlands
\and
Jodrell Bank Centre for Astrophysics, The University of Manchester, Alan Turing Building, Manchester M13 9PL, UK
\and
Instituto de Radioastronom\'{\i}a y Astrof\'{\i}sica, Universidad Nacional Aut\'onoma de M\'exico, Apdo.\ Postal 72-3 (Xangari), Morelia, Michoac\'an 58089, M\'exico
\and
Department of Astrophysics/IMAPP, Radboud University, PO Box 9010, 6500 GL Nijmegen, The Netherlands
\and
ALLEGRO/Leiden Observatory, Leiden University, PO Box 9513, NL-2300 RA Leiden, The Netherlands
\and
School of Physics \& Astronomy, E.\ C.\ Stoner Building, The University of Leeds, Leeds LS2 9JT, UK
\and
UK Astronomy Technology Centre, Royal Observatory Edinburgh, Blackford Hill, Edinburgh EH9 3HJ, UK
\and
Institute of Astronomy and Astrophysics, University of T\"ubingen, Auf der Morgenstelle 10, D-72076 T\"ubingen, Germany
\and
Instituto de Astrof\'{\i}sica e Ci\^encias do Espa\c{c}o, Universidade do Porto, CAUP, Rua das Estrelas, 4150-762, Porto, Portugal
\and
Centre for Astrophysics, University of Hertfordshire, College Lane, Hatfield, AL10 9AB, UK
\and
Leiden Observatory, Leiden University, PO Box 9513, 2300 RA Leiden, The Netherlands 
\and
Max-Planck-Institut f\"{u}r Astrophysik, Karl-Schwarzschild-Str.\ 1, D-85748 Garching, Germany
\and
European Southern Observatory, Karl-Schwarzschild-Str.\ 2, D-85748 Garching, Germany
}
\offprints{M.\ T.\ Beltr\'an, \email{mbeltran@arcetri.astro.it}}
\date{Received date; accepted date}

\titlerunning{Infall and rotation in G31.41+0.31}
\authorrunning{Beltr\'an et al.}

\abstract
{As part of our effort to search for circumstellar disks around high-mass stellar objects, we observed the well-known core G31.41+0.31 with ALMA at 1.4\,mm with an angular resolution of $\sim$$0\farcs22$ ($\sim$1700\,au). The dust continuum emission has been resolved into two cores namely Main and NE. The Main core, which has the stronger emission and  is the more chemically rich, has a diameter of $\sim$5300\,au, and is associated with two free-free continuum sources. The Main core looks featureless and homogeneous in dust continuum emission and does not present any hint of fragmentation. Each transition of \MCN\ and \MF, both ground and vibrationally excited, as well as those of \MCN\ isotopologues, shows a clear velocity gradient along the NE--SW direction, with velocity linearly increasing with distance from the center, consistent with solid-body rotation. However,  when comparing the velocity field of transitions with different upper level energies, the rotation velocity increases with increasing energy of the transition, which  suggests that the rotation speeds up towards the center. Spectral lines towards the dust continuum peak show an inverse P-Cygni profile that supports the existence of infall in the core. The infall velocity increases with the energy of the transition suggesting that the infall is accelerating towards the center of the core, consistent with gravitational collapse. Despite the monolithic appearance of the Main core, the presence of red-shifted absorption, the existence of two embedded free-free sources at the center, and the rotational spin-up are consistent with an unstable core undergoing fragmentation with infall and differential rotation due to conservation of angular momentum. Therefore,  the most likely explanation for the monolithic morphology is that the large opacity of the dust emission prevents the detection of any inhomogeneity in the core.}

\maketitle

\section{Introduction}

The formation process of high-mass stars has puzzled the astrophysical community for decades because of the apparent stellar mass limit for spherical accretion. Beyond this limit, theory predicted that it is impossible to continue accreting material because the stellar wind and the radiation pressure from the newly-formed early-type star would stop the infall (e.g., Kahn~\cite{khan74}; Yorke \& Kr\"ugel~\cite{yorke77}; Wolfire \& Casinelli~\cite{wolfire87}). Different theoretical scenarios have proposed non-spherical accretion as a possible solution for the formation of OB-type stars (Nakano~\cite{nakano89}; Jijina \& Adams~\cite{jijina96}), and in recent years, all models appear to have converged to a disk-mediated accretion scenario (e.g., Krumholz et al.~\cite{krumholz09}; Kuiper et al.~\cite{kuiper10}, \cite{kuiper11}; Peters et al.~\cite{peters10a}; Kuiper \& Yorke~\cite{kuiper13}; Klassen et al.~\cite{klassen16}). Competing theories that propose very different high-mass star-formation mechanisms, such as models suggesting that massive star-formation is initiated by the monolithic collapse of a turbulent molecular core (McKee \& Tan~\cite{mckee02}), or those based
on competitive accretion (Bonnell \& Bate 2006),  all predict the existence of circumstellar accretion disks through which the material is channeled onto the forming star. High-angular resolution observations have confirmed these theoretical prediction and, in recent years, especially since the advent of ALMA, circumstellar disks around B-type stars, which have luminosities $<10^5\,L_\odot$, have been discovered.  In some cases,  it has been proposed that these structures could be true accretion disks undergoing Keplerian rotation  (see the review by Beltr\'an \& de Wit~\cite{beltran16}).  The detection of Keplerian disk candidates has been recently extended up to more massive late O-type stars, with luminosities of $\sim$$10^5\,L_\odot$ (Johnston et al.~\cite{johnston15}; Ilee et al.~\cite{ilee16}). 

From an observational point of view, the natural follow-up to these findings is to search for disks around early O-type stars, with spectral types earlier than O6--O7, and luminosities $>10^5\,L_\odot$, to test if stars of all masses could form through disk-mediated accretion. With this in mind, we carried out a survey with ALMA of six O-type star-forming regions associated with known hot molecular cores (HMCs) looking for circumstellar disks. The first analysis of these observations has revealed different degrees of fragmentation of the cores and the presence of three Keplerian disk candidates (Cesaroni et al.~\cite{cesa17}). One of the HMCs of this survey is  G31.41+0.31 (hereafter G31), for which our ALMA observations have revealed no signs of fragmentation and a velocity gradient consistent with rotation.  

G31 is a well known HMC with a luminosity $\ga$$10^5\,L_\odot$ (Cesaroni et al.~\cite{cesa94}) located at a kinematic distance of 7.9~kpc  (Churchwell et al.~\cite{churchwell90}). 
Alternatively, G31 might be associated with the W43--Main cloud complex, as suggested by Nguyen-Luong et al.~(\cite{nguyen11}). For W43--Main, two distance estimates exist which are based on VLBI observations of maser parallax. Reid et al.~(\cite{reid14}) report a distance of 4.9\,kpc to the W43--Main core, while Zhang et al.~(\cite{zhang14}) report distances to 5 maser spots ranging from 4.27 to 6.21\,kpc. Given the uncertain association of G31 with W43--Main and the uncertainty in the distance to W43--Main itself, we adopt the kinematic distance of 7.9\,kpc for G31 in this paper. However, one should keep in mind that the real distance may be substantially less ($\sim$5\,kpc) which would imply a $\sim$2.5 times lower luminosity.

The G31 hot core has a size of $\sim$1$''$ 
($\sim$8000\,au) and its  mass is $\gtrsim500\,M_\odot$ (Beltr\'an et al.~\cite{beltran04}; Girart et al.~\cite{girart09}; Cesaroni et al.~\cite{cesa11}). The core is $\sim$5$''$ away from an ultracompact (UC) \HII region 
and overlaps in projection a diffuse halo of free-free emission,  possibly associated with the \UC\ region itself.  Given the intensity of the molecular line and continuum emission, the core has been extensively studied with both single-dish and interferometric observations (e.g., Cesaroni et al.~\cite{cesa94}; Olmi et al.~\cite{olmi96}; Beltr\'an et al.~\cite{beltran04},  \cite{beltran05}; Girart et al.~\cite{girart09}; Cesaroni et al.~\cite{cesa11}; Mayen-Gijon et al.~\cite{mayen14}). These observations, mostly at millimeter wavelengths, have confirmed that the core is very chemically rich, presenting prominent emission in a large number
of complex organic molecules, some of them of pre-biotic importance (e.g., Beltr\'an et al.~\cite{beltran09}; Rivilla et al.~\cite{rivilla17}). The molecular emission has been successfully used to trace the gas distribution and kinematics of the dense, hot gas and estimate important physical parameters (e.g., temperature and density: Beltr\'an et al.~\cite{beltran05}). Molecular line observations have also revealed the existence of a striking velocity gradient (centered at an LSR velocity of $\sim$96.5\,\kms) across the core in the northeast--southwest (NE--SW) direction (Beltr\'an et al.~\cite{beltran05}; Cesaroni et al.~\cite{cesa11}). This has been interpreted as rotational motion around embedded massive stars, as suggested by the detection of two point-like ($<$0\farcs07) free-free continuum sources close to the core center (Cesaroni et al.~\cite{cesa10}). Polarization measurements by Girart et al.~(\cite{girart09}) have revealed an hour-glass shaped magnetic field of $\sim$10~mG, with the symmetry axis oriented perpendicular to the velocity gradient. Moreover, the same authors have detected inverse P-Cygni profiles in a few molecular lines, indicating that infall motion is present (see also Mayen-Gijon et al.~\cite{mayen14}). Wyrowski et al.~(\cite{wyrowski12}) have also detected red-shifted absorption in ammonia with SOFIA. All these features are consistent with a scenario in which the core is contracting and rotating about the direction of the magnetic field lines.

\begin{figure}
\centerline{\includegraphics[angle=0,width=7.5cm,angle=90]{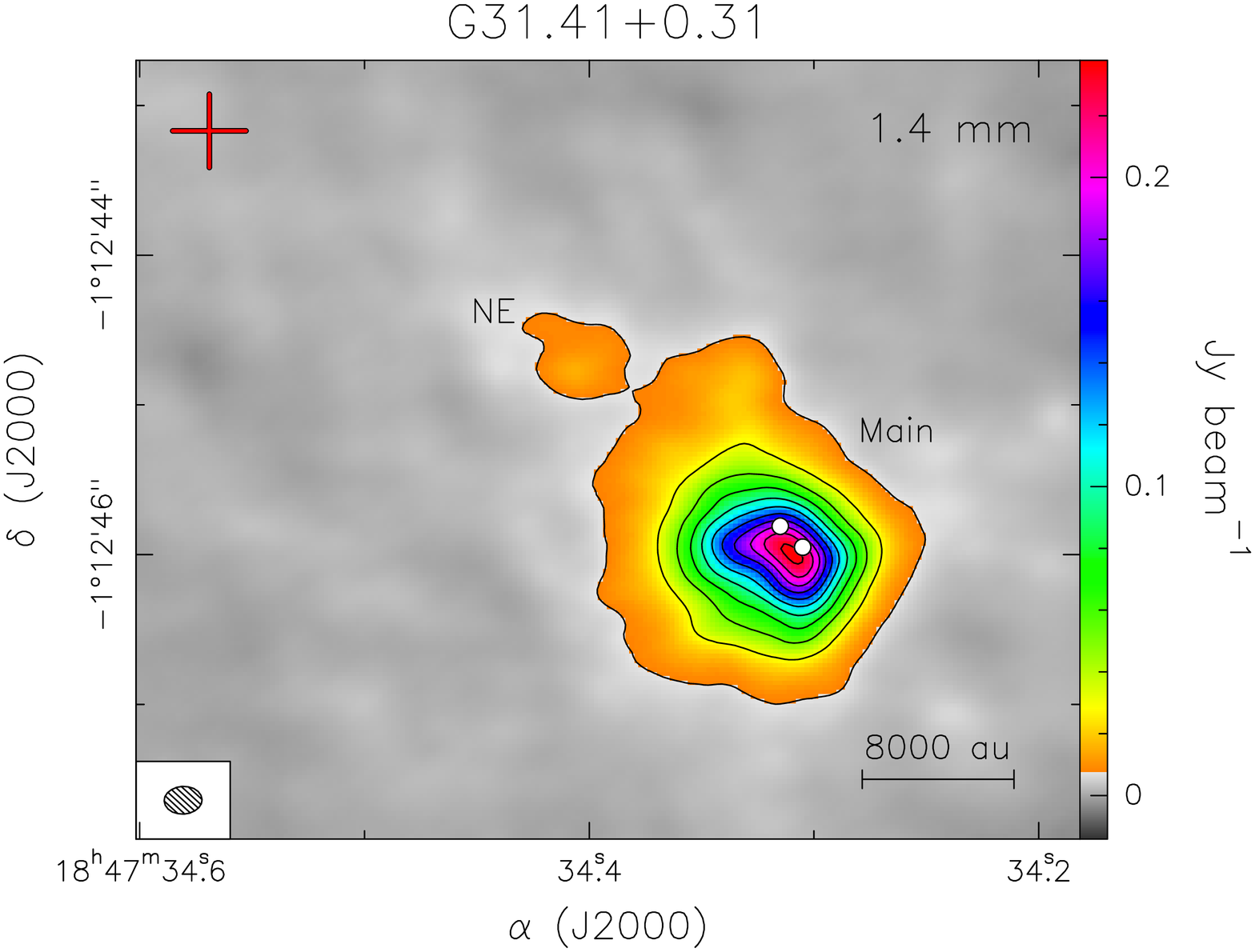}}
\caption{ALMA map of the 1.4\,mm continuum emission from the G31 HMC.  The contours range from 7.5 (5$\sigma$) to 
232.5\,mJy\,beam$^{-1}$ in steps of 22.5\,mJy\,beam$^{-1}$ (15$\sigma$). The white dots mark the position of two unresolved free-free continuum sources detected at wavelengths between 3.6\,cm and 7\,mm by Cesaroni et al.~(\cite{cesa10}). The red cross indicates the position of the \UC\ region.
The synthesized beam is shown in the lower left-hand corner. }
\label{fig-cont}
\end{figure}

In this work, we analyze in more detail the ALMA observations of G31 previously presented by Cesaroni et al.~(\cite{cesa17}). These observations have an angular resolution of $\sim$$0\farcs22$ ($\sim$1700\,au at the distance of the source), which is $\gtrsim$4 times higher than that of previous (sub)millimeter observations (e.g., Beltr\'an et al.~\cite{beltran05}; Girart et al.~\cite{girart09}; Cesaroni et al.~\cite{cesa11}) and therefore allow us to study the morphology and kinematics of this core with unprecedented detail. In particular, we investigate whether the compact appearance of G31, with no hint of fragmentation despite its large mass, is real.   We study the physical conditions of the core and of the molecular outflows associated with it. Regarding the kinematics, by analyzing the velocity gradient, we discuss what type of rotation the core could be undergoing, and by analyzing the red-shifted absorption detected towards the center of the core, we discuss its possible collapse.

\section{Observations}

Interferometric observations of \g31 were carried out with ALMA in Cycle 2 in
July and September 2015 as part of project 2013.1.00489.S. (P.I.: R.\ Cesaroni). The observations were performed 
in Band 6 with the array in an extended configuration.  The digital correlator was configured in thirteen spectral
windows (SPW) that cover among other lines, SiO\,(5--4), \MCN\,(12--11), \MCNII\,(12--11), \MCN\,$v_8=1$\,(12--11), and \MCNI\,(13--12). We refer to  
Cesaroni et al.~(\cite{cesa17}) for detailed information on the observations, and, in particular, on the continuum subtraction.  

The phase reference center of the observations is 
$\alpha$(J2000)=18$^{\rm h}$47$^{\rm m}$34$\fs$315,
$\delta$(J2000)=$-01^\circ$12$'$45$\farcs$90.  The position uncertainty is $<$$0\farcs2$. The
data were calibrated and imaged using the {\sc CASA}\footnote{The {\sc CASA} package is
available at \url{http://casa.nrao.edu/}} software package (McMullin et al.~\cite{mcmullin07}). Further imaging and
analysis were done with the {\sc GILDAS}\footnote{The {\sc GILDAS} package is available at
\url{http://www.iram.fr/IRAMFR/GILDAS}} software package.  A total of 13 individual data cubes were created using the CLEAN task and the ROBUST parameter of  Briggs~(\cite{briggs95}) set equal to 0.5. The continuum was determined from the broadest spectral window, centered at 218 GHz, using the {\sc STATCONT}\footnote{\url{http://www.astro.uni-koeln.de/~sanchez/statcont}} algorithm (S\'anchez-Monge et al.~\cite{sanchez-monge18}).  The resulting synthesized CLEANed beam of the maps is $0\farcs25\times0\farcs19$ for the continuum, and ranges from $0\farcs25\times0\farcs19$ to $0\farcs27\times0\farcs20$ for the lines analyzed in this work. These angular resolutions correspond to spatial scales of $\sim$1700\,au at the distance of the source.  The rms noise of the maps is 1.5\,mJy\,beam$^{-1}$ for the continuum and $\sim$1.3\,mJy\,beam$^{-1}$ per channel of 2.7~\kms, $\sim$1.6\,mJy\,beam$^{-1}$ per channel of 0.66\,\kms, and $\sim$2.1\,mJy\,beam$^{-1}$ per channel of 0.33\,\kms\ for the line maps. 
  The total bandwidth of the spectral window used to estimate the continuum is 1875\,MHz and the spectral resolution is $\sim$1.95\,MHz.  The number of channels with emission considered to be continuum by STATCONT ranges from 144, which corresponds to a bandwidth of 280\,MHz, at line-poor positions, to 10, which corresponds to a bandwidth of 17\,MHz, at line-rich positions.  However, the rms noise of the continuum map would indicate an effective bandwidth of $\sim$2\,MHz if the noise were just thermal noise. This along  with the fact that the continuum source is very strong leads us to conclude that the high rms noise of such map is due to dynamic range problems and not  to the STATCONT method used to subtract the continuum.

\begin{table*}
\caption[] {Positions, flux densities, and diameters of the cores$^{a}$.}
\label{table-cont}
\begin{tabular}{lcccccccc}
\hline
&\multicolumn{2}{c}{Peak position}
&&&&\multicolumn{3}{c}{Source diameter}
\\
 \cline{2-3} 
 \cline{7-9} 
&\multicolumn{1}{c}{$\alpha({\rm J2000})$} &
\multicolumn{1}{c}{$\delta({\rm J2000})$} &
\multicolumn{1}{c}{$T_{\rm B}^b$} &
\multicolumn{1}{c}{$I^{{\rm peak}, c}_\nu$} &
\multicolumn{1}{c}{$S_\nu^c$} &
\multicolumn{1}{c}{FWHM$^{d}$} &
\multicolumn{1}{c}{$\theta_s$$^{e}$} &
\multicolumn{1}{c}{$D_s$$^{e}$}  
\\
\multicolumn{1}{c}{Core} &
\multicolumn{1}{c}{h m s}&
\multicolumn{1}{c}{$\degr$ $\arcmin$ $\arcsec$} &
\multicolumn{1}{c}{(K)} & 
\multicolumn{1}{c}{(Jy/beam)} & 
\multicolumn{1}{c}{(Jy)} &
\multicolumn{1}{c}{(arcsec)} &
\multicolumn{1}{c}{(arcsec)} &
\multicolumn{1}{c}{(au)} 
\\
\hline
G31-Main    &18 47 34.309 &$-$01 12 45.99  &132$\pm$1 &0.238$\pm$0.002    &3.10$\pm$0.05  &0.70$\pm$0.04 &0.67$\pm$0.04  &5300$\pm$300  \\
G31-NE       &18 47 34.407 &$-$01 12 44.79  &8$\pm$1    &0.015$\pm$0.002     &0.052$\pm$0.002  &0.61$\pm$0.04 &0.56$\pm$0.04  &4500$\pm$300  \\
\hline
\end{tabular}

 $^a$ Derived from the continuum dust emission. \\
 $^b$ The conversion factor from Jy/beam to K is 555.5. \\
  $^c$ Peak intensity and integrated flux density corrected for primary beam
  response.  \\
  $^d$ FWHM = $2\sqrt{A/\pi}$, where $A$ is the area inside the contour at half maximum. \\
  $^e$ Deconvolved diameter, calculated assuming that the cores are symmetric Gaussians, using the expression $\theta_s=\sqrt{{\rm FWHM}^2-{\rm HPBW}^2}$, where HPBW is the half-power width of the synthesized beam.  The distance to G31 is assumed to be 7.9\,kpc.\\ 
\end{table*}

\section{Results}
\subsection{Continuum emission}
\label{cont-em}

Figure~\ref{fig-cont} shows the continuum emission map of the G31 hot core at 1.4\,mm (217\,GHz). The dust emission is resolved into two cores, a strong and large one located at the center of the map that we call Main, and a smaller and weaker one located to the northeast of the Main core, which we call NE.  The \UC\ region, located to the northeast  of the HMC, is not detected in the continuum emission at 1.4\,mm.  The position, flux, and diameter of the cores are given in Table~\ref{table-cont}.  The angular diameter has been computed as the diameter of the circle whose area equals that inside the  50\% contour level.

The dust continuum emission of the Main core  peaks close to the position of one of the two unresolved free-free continuum sources detected from 3.6\,cm to 7\,mm by Cesaroni et al.~(\cite{cesa10}). As seen in Fig.~\ref{fig-cont}, this core appears quite round, uniform, and compact, with no hint of fragmentation, despite the fact that the angular resolution of the observations is high enough to properly resolve the emission. In fact, the deconvolved half-power diameter of the core is 
$\sim$5300\,au, much larger than the spatial resolution of the observations ($\sim$1700\,au). The deconvolved diameter of the source at the  5$\sigma$ emission level is $\sim$17000\,au, 10 times larger than the synthesized beam. By fitting elliptical Gaussians to the dust emission map, we measured a deconvolved
size of the source of $0\farcs9\times0\farcs7$ (7000$\times$5500\,au) with a position angle (PA) of $\sim$63$^\circ$.

The NE core is much weaker than the Main core. Its integrated flux density is $\sim$60 times lower and its peak intensity $\sim$16 times lower.  This source is much smaller than the Main core and has a 5$\sigma$ emission level deconvolved diameter of $\sim$4500\,au, which coincides with the deconvolved diameter of the 50\% contour level. 

\subsection{Line emission}

Figures~\ref{fig-spectra} and \ref{fig-spectra-NE} show the spectra observed in the 13  spectral windows 
toward the Main and the NE core, respectively. The spectra were obtained by integrating over the $5\sigma$ contour level area of the 
dust continuum emission.  Clearly, the Main core is the most chemically rich, since the corresponding spectrum
displays a forest of molecular lines. For this reason, we postpone a detailed analysis of the molecular line emission
toward this core to a future study (Rivilla et al., in prep.). In the present work, we focus on i) two well-known hot core (high-density) tracers:  
\MCN\ (ground state and vibrationally excited) and its isotopologues \MCNII\ and \MCNI, and \MF\ (ground state and vibrationally excited), and ii) a typical outflow tracer: SiO. Occasionally, we also discuss the emission of the lower density tracer H$_2$CO. The different transitions of these species are indicated in Figs.~\ref{fig-spectra} and \ref{fig-spectra-NE}.  The typical cloud and/or molecular outflow tracers C$^{18}$O and $^{13}$CO are also covered by our spectral setup but  their emission is so heavily filtered out by the interferometer that it has been impossible to use these lines for our study.

\begin{figure*}
\centerline{\includegraphics[angle=0,width=16cm,angle=0]{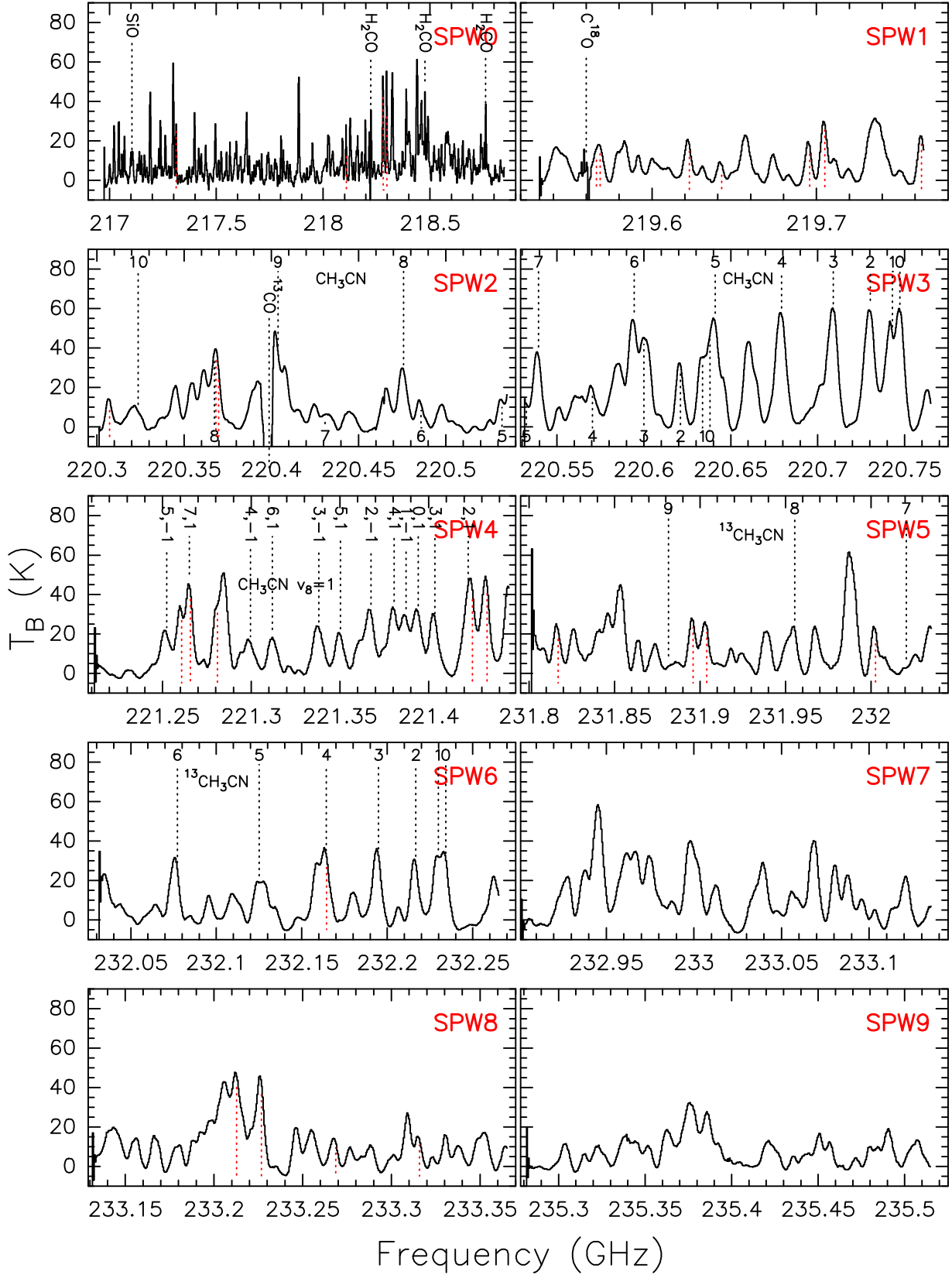}}
\caption{Continuum-subtracted spectra obtained by integrating the emission over the 5$\sigma$ contour level area of the continuum emission of the Main core in G31. The spectra shown are spread over the
whole frequency range of the ALMA observations. Different $K$ numbers are marked with dotted lines in the upper 
part of each spectra in the case of CH$_3$CN (SPW2 and 3),  vibrationally excited CH$_3$CN  (SPW4), and \MCNI\  (SPW5 and 6), and in the lower part in the case of \MCNII\ (SPW2 and 3). The red dotted lines in the lower part of each spectra indicate \MF\  $v=0$ and $v_t=1$ transitions.
The different spectral windows (SPW) are indicated in red in the top right of each panel.}
\label{fig-spectra}
\end{figure*}

\begin{figure*}
\addtocounter{figure}{-1}
\centerline{\includegraphics[angle=0,width=16cm,angle=0]{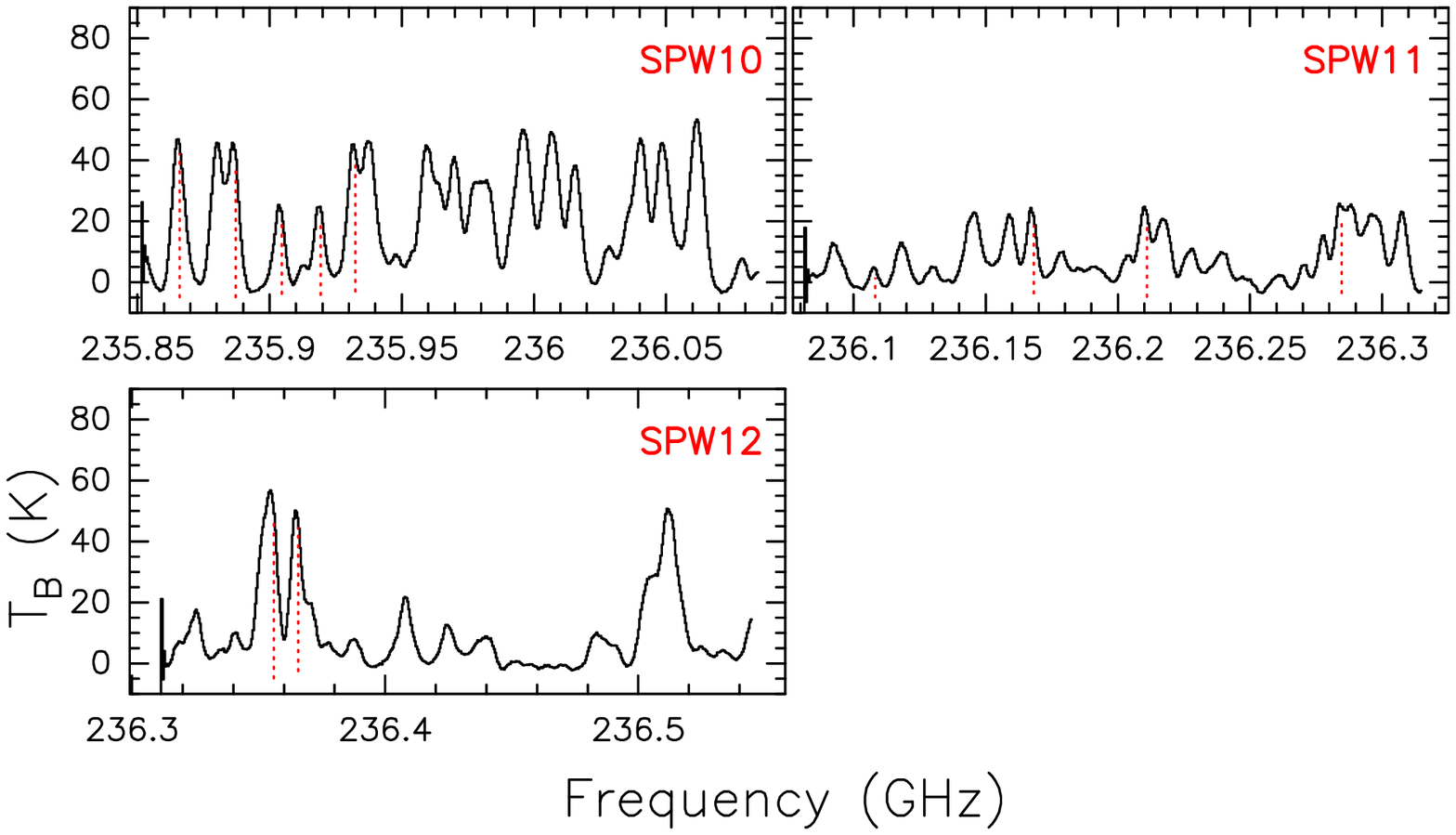}}
\caption{Continued}
\end{figure*}

\begin{figure*}
\centerline{\includegraphics[angle=0,width=16cm,angle=0]{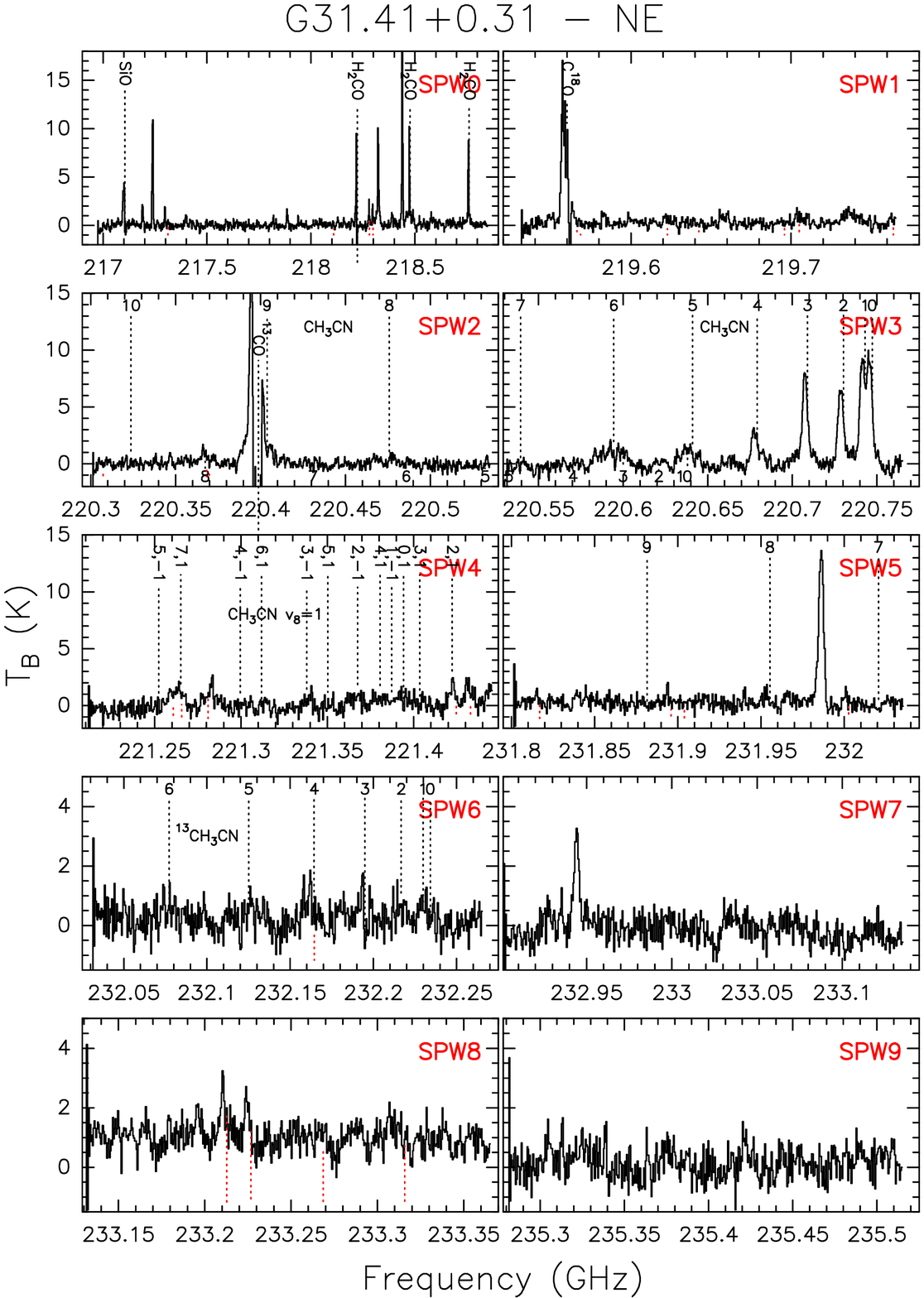}}
\caption{Same as Fig.~\ref{fig-spectra} for the NE source.}
\label{fig-spectra-NE}
\end{figure*}

\begin{figure*}
\addtocounter{figure}{-1}
\centerline{\includegraphics[angle=0,width=16cm,angle=0]{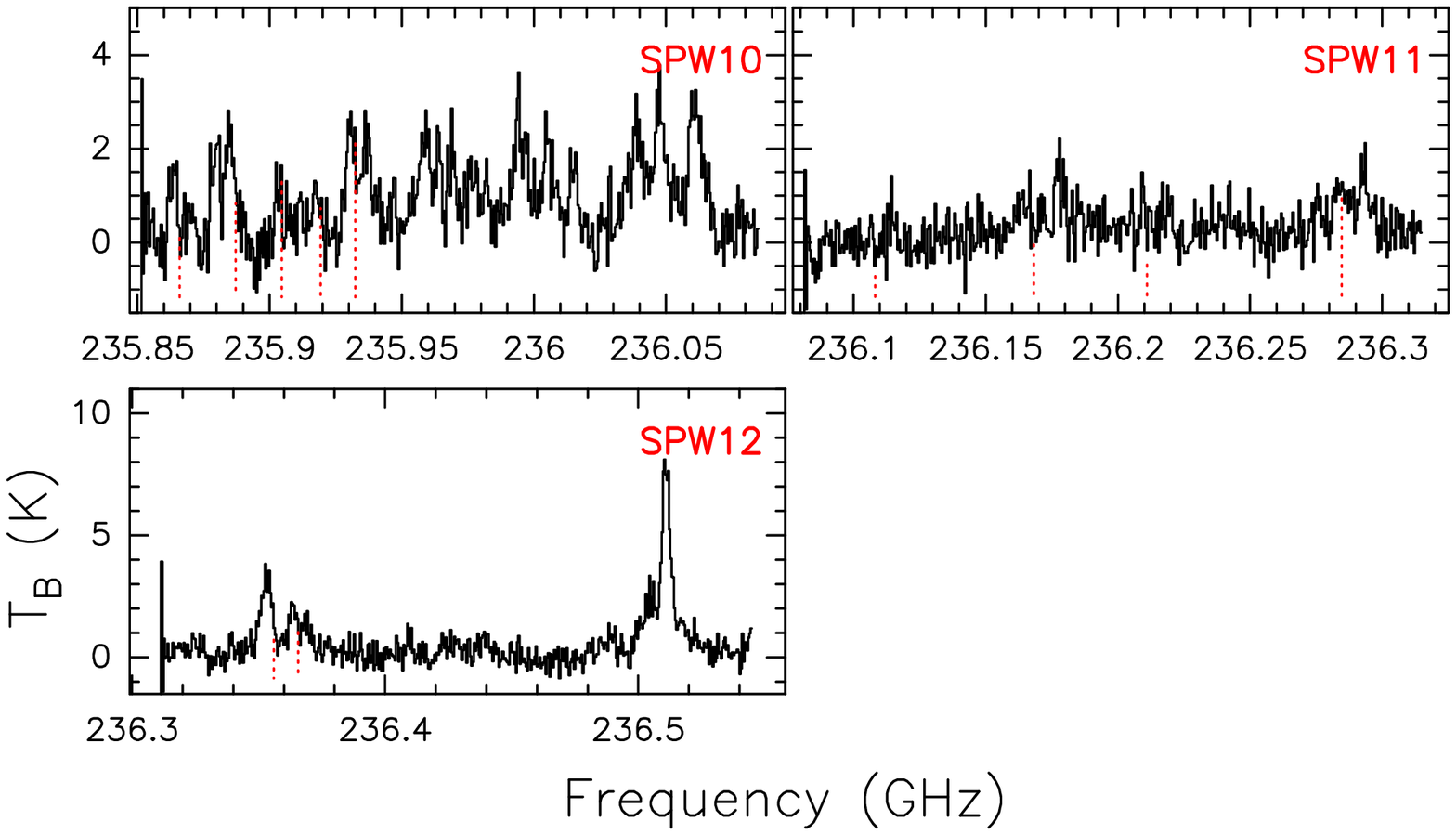}}
\caption{Continued}
\end{figure*}

\begin{figure*}
\centerline{\includegraphics[angle=0,width=16cm,angle=90]{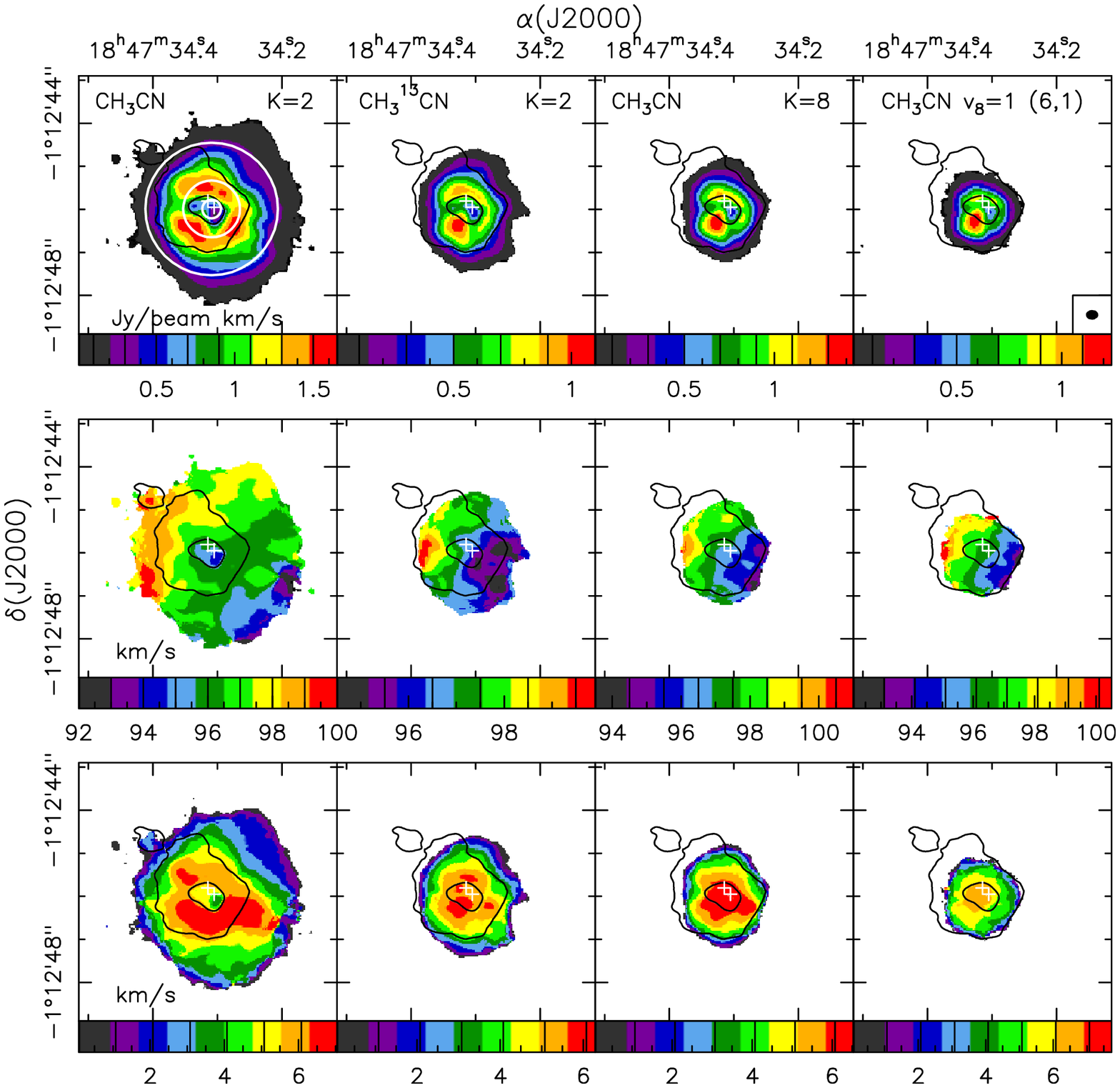}}
\caption{{\it Top panels}: overlay of the 217\,GHz continuum emission ({\it black contours}) on the integrated intensity (moment 0) map ({\it colors}) of \MCN\
$K=2$, \MCNII\ $K=2$, \MCN\ $K=8$, and \MCN\  $K, l=(6, 1)$ $v_8=1$ (12--11). The contours correspond to the 5$\sigma$ and 50\% levels. The concentric white circles in the first panel have radii of $0\farcs22$, $0\farcs66$, and $1\farcs54$, (see Sects.~\ref{fitting} and \ref{t-gradient}). {\it Middle panels}: line velocity (moment 1) maps for the same molecular species. {\it  Bottom panels}: velocity dispersion (moment 2) maps for the same species. The white crosses indicate the positions of the two compact free-free continuum sources detected by Cesaroni et al.~(\cite{cesa10}). The synthesized beam is shown in the lower right-hand corner of the top right panel. }
\label{fig-gradients}
\end{figure*}

\begin{figure*}
\centerline{\includegraphics[angle=0,width=16cm,angle=90]{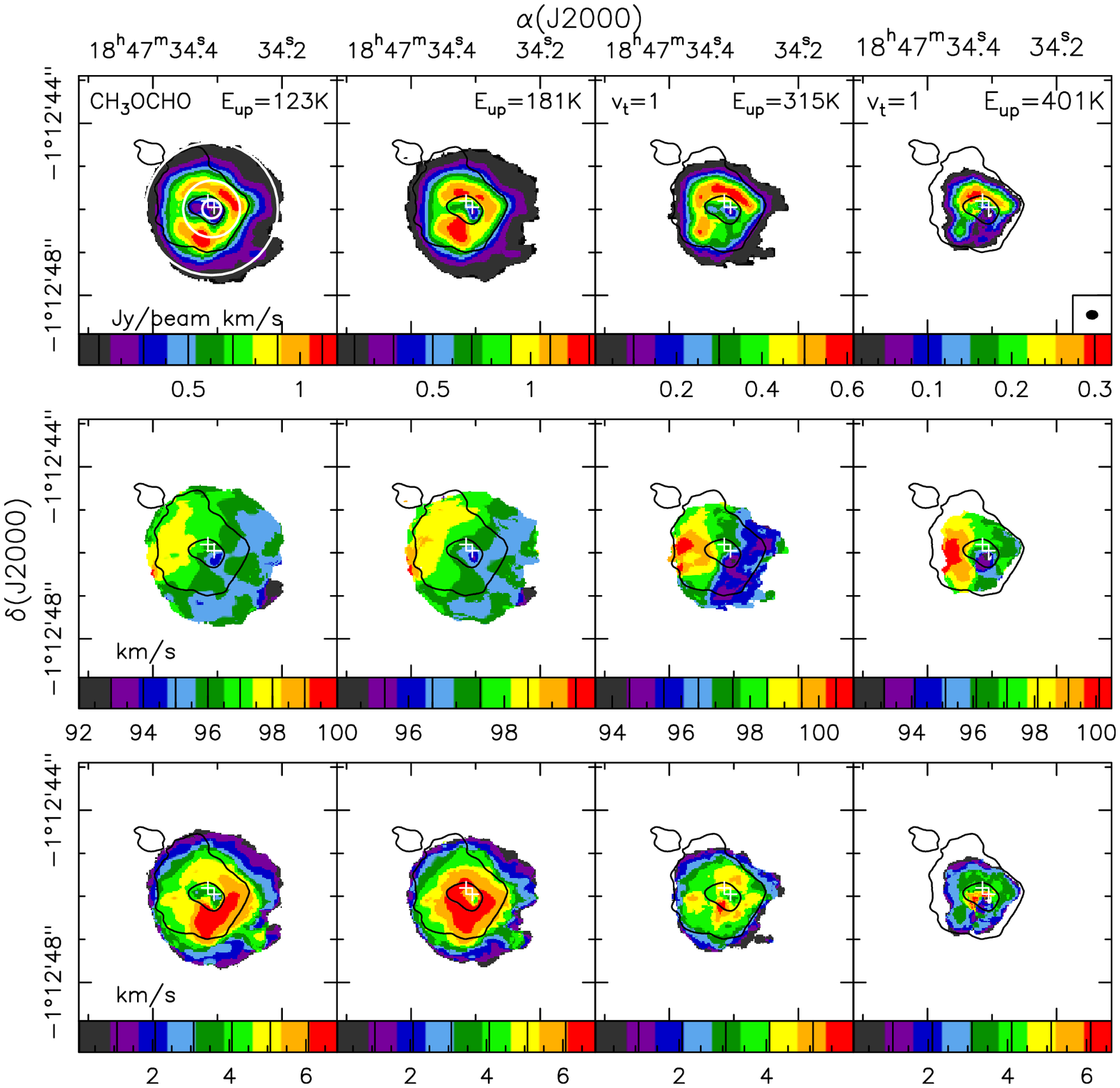}}
\caption{Same as Fig.~\ref{fig-gradients} for \MF\ (19$_{4, 16}-18_{4, 15}$)A, (18$_{11, 7}-17_{11, 6}$)A, (20$_{2, 18}-19_{2, 17}$)A v$_t$=1, and  (18$_{13, 6}-17_{13, 5}$)E v$_t$=1. The upper level energies $E_{\rm up}$ of each transition are indicated in the upper right-hand corner of the top panels. }
\label{fig-gradients-MF}
\end{figure*}

\subsubsection{\MCN\ and isotopologues, \MF, and H$_2$CO}
\label{high-tracers}

The different $K$ transitions of \MCN\ (including the vibrationally excited transitions) and the two isotopologues \MCNII, and \MCNI\ cover a broad range of excitation conditions, with upper level energies ranging from $\sim$70 K to $\sim$860 K. On the other hand, our setup covers almost 2000 transitions of \MF, including both ground and vibrationally excited, with upper level energies of $\sim$20--1700\,K. In our data, many of these transitions are strongly blended with other species (and some are simply too faint to be detected). In the end we have analyzed only those transitions of \MF\ that are unblended or slightly blended with other species, which in total number are $\sim$40 transitions.  These transitions cover a range of energies of $\sim$100--440\,K. 

\begin{figure*}
\centerline{\includegraphics[angle=0,width=16cm,angle=0]{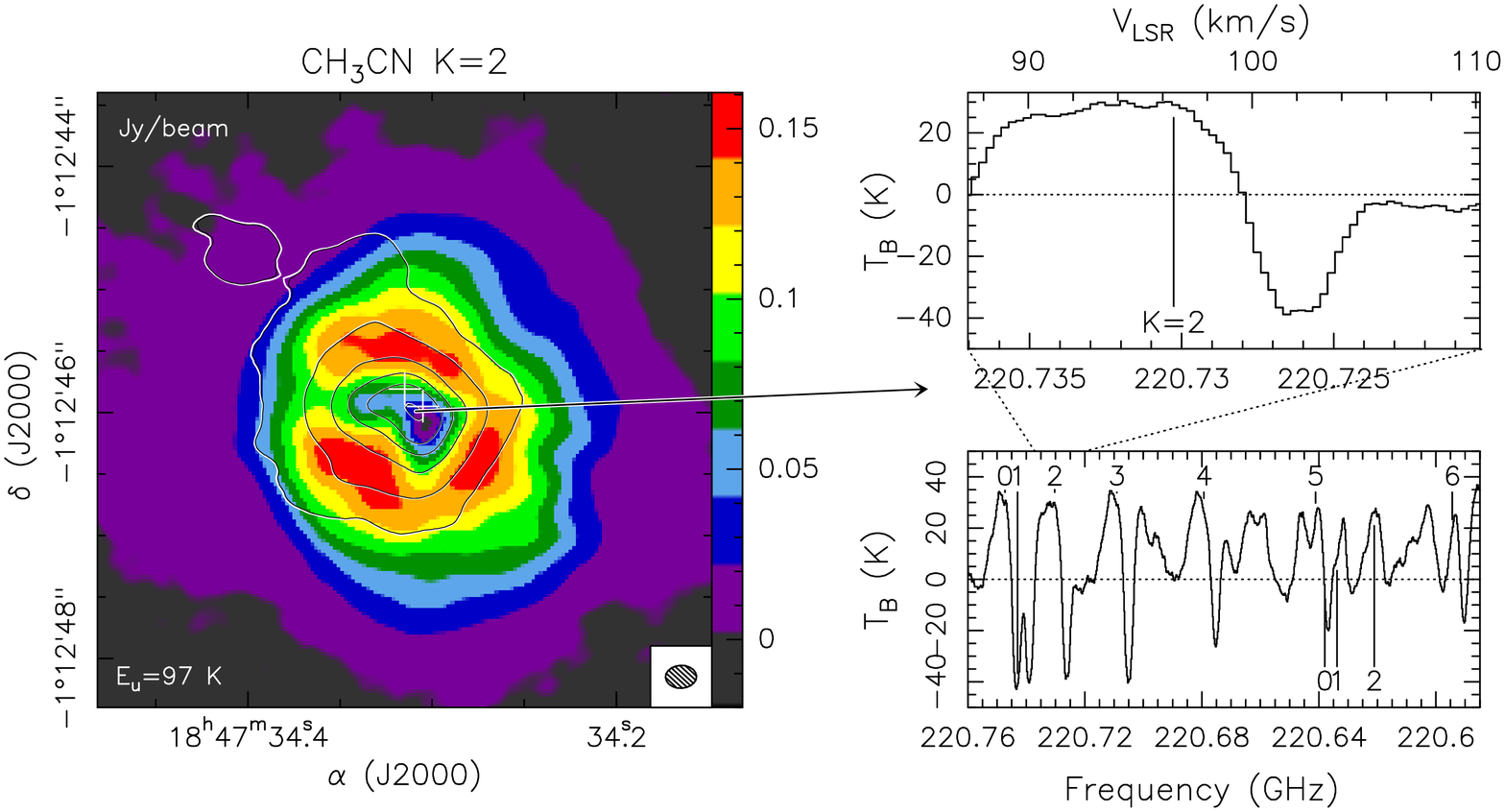}}
\caption{{\it Left panel}: Dust continuum emission map ({\it contours}) overlaid on the  map of the \MCN\ (12--11)  $K$=2 emission averaged over the velocity range 92 to 102\,\kms\ ({\it colors}). Contours are 7.5, 30, 75, 120, 165, and 232.5\,\mjy. The white crosses indicate the positions of the two compact free-free continuum sources detected by Cesaroni et al.~(\cite{cesa10}).  The synthesized beam is shown in the lower right-hand corner. {\it  Right panels}: Spectra of the \MCN\ $K$=0 to 6 and  \MCNII\ $K$=0 to 2 towards the absorption dip, that is, towards the dust continuum peak. The  {\it upper right panel} zooms in on the  \MCN\ $K$=2 spectrum.}
\label{fig-abs}
\end{figure*}

These high-density tracers have allowed us to study the emission of the hot core at different excitation conditions, likely tracing material at different depths in the core because one expects the highest energy transitions to be optically thinner and trace material closer to the central (proto)star(s).  As shown in Figs.~\ref{fig-gradients} and \ref{fig-gradients-MF}, the integrated emission of all species  (methyl cyanide and isotopologues, and methyl formate) traces a ring-like structure with the dust continuum emission peak and the two free-free sources located at the central  dip.  This morphology is better seen in the lowest energy transitions such as  \MCN\, $K=2$ ($E_{\rm up}$=97\,K) or \MF\ (19$_{4, 16}$--18$_{4, 15}$)A ($E_{\rm up}$=123\,K), and is still visible in transitions with energies as high as 778\,K (e.g., \MCN\, $K,l=6, 1$ $v_8=1$). For this vibrationally excited transition, although the emission still decreases towards the central region, the dip is less pronounced. This ring-like morphology in \MCN\ had already been observed by 
Beltr\'an et al.~(\cite{beltran05}) and Cesaroni et al.~(\cite{cesa11}) with the IRAM PdBI and SMA interferometers with an angular resolution of $\sim$1$''$. These studies suggest that the decrease of the \MCN\ emission might be produced by self-absorption due to the high opacities in the central region of the core, combined with a temperature gradient, although none of the \MCN\ lines showed evidence of self-absorption or inverse P-Cygni profiles. As one can see in Fig.~\ref{fig-abs}, the situation changes when the core is observed with an angular resolution of $\sim$$0\farcs2$ because the \MCN\ emission towards the dust continuum emission peak clearly shows red-shifted absorption. Inverse P-Cygni profiles had been previously detected by Girart et al.~(\cite{girart09}) in low energy lines such as C$^{34}$S\,(7--6), and although less evident, also in  CH$_3$OH and isotopologues  with SMA 1$''$ angular resolution observations. This is the first time that the red-shifted absorption in \g31 has been observed in \MCN\ and isotopologues. 

\begin{figure}
\centerline{\includegraphics[angle=0,width=9cm,angle=0]{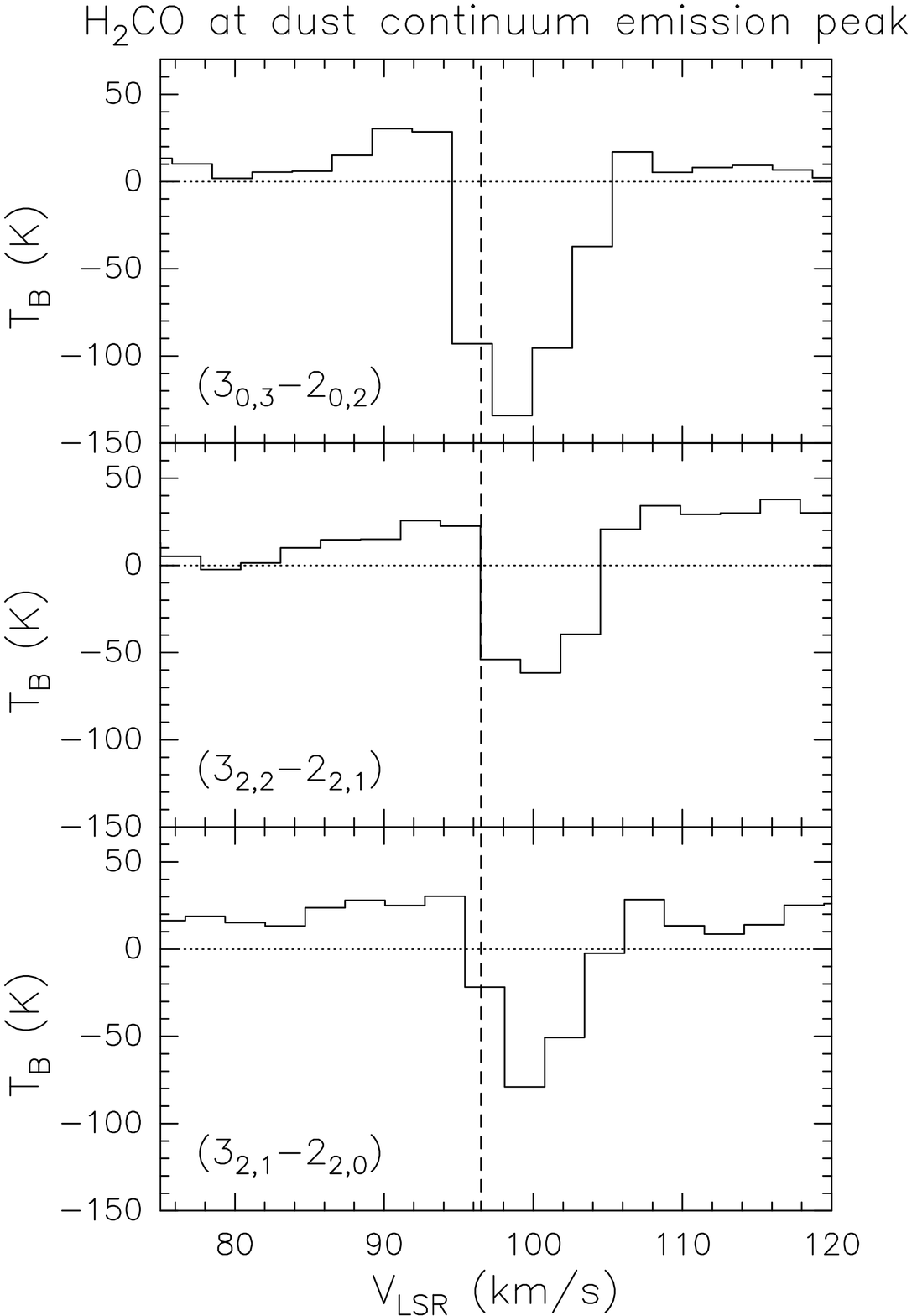}}
\caption{H$_2$CO (3$_{0,3}$--2$_{0,2}$), (3$_{2,2}$--2$_{2,1}$), and (3$_{2,1}$--2$_{2,0}$) spectra at the dust continuum emission peak. The vertical dashed line indicates the  systemic LSR velocity of 96.5\,\kms.}
\label{spectra-h2co}
\end{figure}

The red-shifted absorption is also clearly detected in H$_2$CO (3$_{0, 3} - 2_{0, 2}$),  (3$_{2, 2} - 2_{2, 1}$), and (3$_{2, 1} - 2_{2, 0}$), which have upper level energies of 21 and 68~K (Fig.~\ref{spectra-h2co}). The absorption is much deeper in H$_2$CO than in CH$_3$CN:  the brightness temperature $T_{\rm B}$ is $-$134\,K for H$_2$CO  (3$_{0, 3} - 2_{0, 2}$) and $-$40\,K for CH$_3$CN $K$=2.

\subsubsection{SiO emission}

\begin{figure*}
\centerline{\includegraphics[angle=0,width=18cm,angle=0]{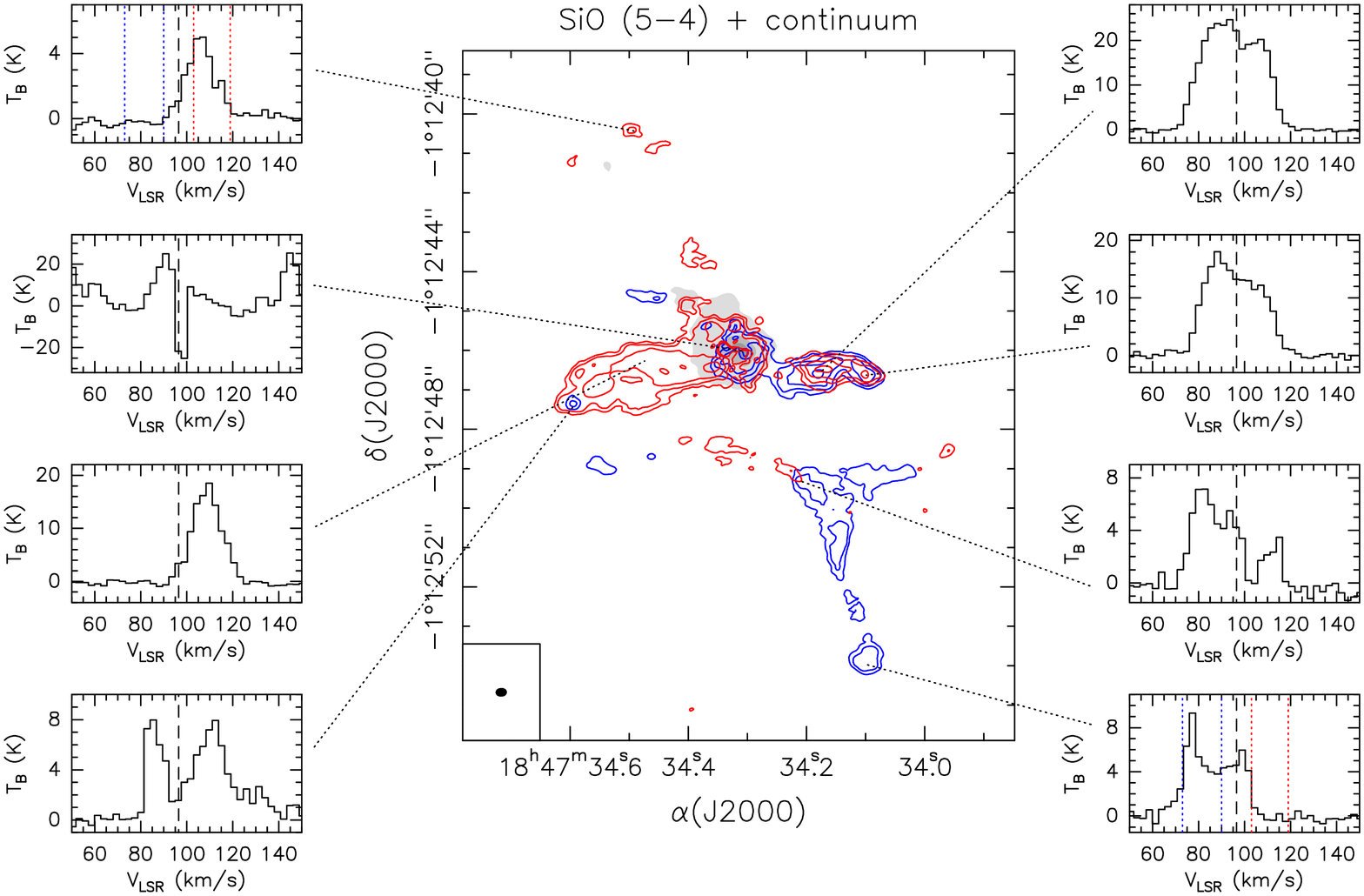}}
\caption{{\it Middle panel}: overlay of the 217\,GHz continuum emission ({\it grayscale}) on the blue-shifted ({\it blue contours}) and red-shifted ({\it red contours}) SiO\,(5--4) averaged emission. The blue-shifted emission has been averaged over the (73, 90)\,\kms\ velocity interval and the red-shifted emission over the (103, 119)\,\kms\ one
and are indicated in blue and red dotted vertical lines in the top left and lower right spectra.  Contour levels are 3, 6, 12,  and 24 times 1$\sigma$, where 1$\sigma$ is 1.1\,mJy\,beam$^{-1}$. Grayscale contours for the continuum emission range from 7.5 to 232.5\,mJy beam$^{-1}$ in steps of 90\,mJy beam$^{-1}$. The synthesized beam is shown in the lower left-hand corner. {\it  Left and right panels}: SiO\,(5--4) spectra towards selected positions in the G31 core. The vertical dashed line indicates the systemic LSR velocity of 96.5\,\kms.}
\label{fig-sio}
\end{figure*}

\begin{figure*}
\centerline{\includegraphics[angle=0,width=14cm,angle=90]{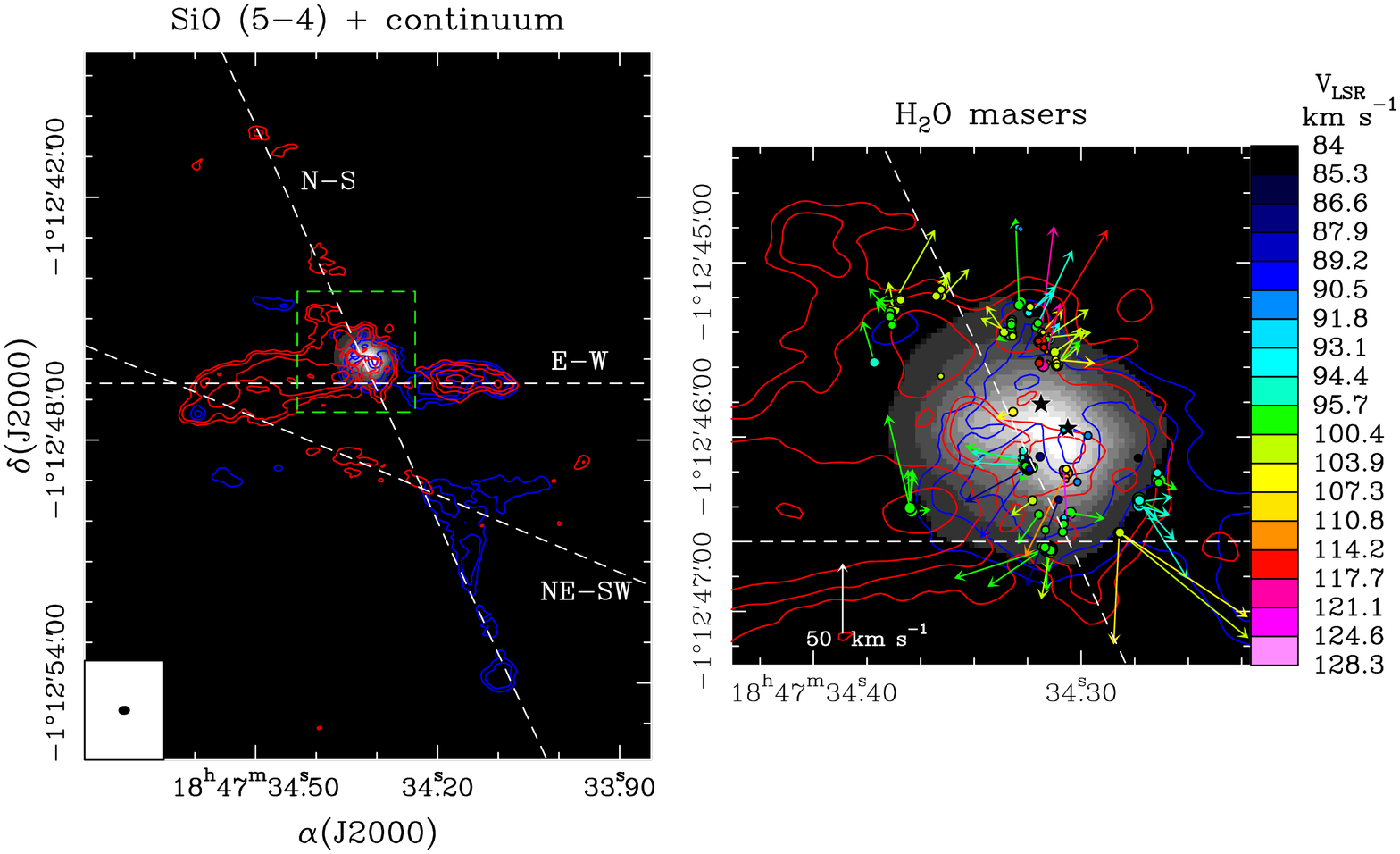}}
\caption{{\it Left panel}:  Same as Fig.~\ref{fig-sio}. The white dashed lines indicate the direction of the three possible outflows in G31, and the green dashed box indicates the area plotted in the close-up panel. Grayscale levels for the continuum are 10, 20, 30, 40, 50, 60, 60, 70, 80, and 90\% of the peak. {\it Right panel}: Close-up of the central region. Colored circles mark the position of H$_2$O masers  while colored vectors indicate the direction and the amplitude of the proper motions (Moscadelli et al.~\cite{mosca13}). The white vector in the
bottom left corner indicates the amplitude scale of proper motions in kilometer per second. The black stars indicate the positions of the two free-free continuum sources detected by Cesaroni et al.~(\cite{cesa10}).}
\label{fig-sio-dir}
\end{figure*}

The molecular outflows associated with \g31 have been studied in CO and $^{13}$CO  by Olmi et al.~(\cite{olmi96}) and  Cesaroni et al.~(\cite{cesa11}). The CO emission is very complex and reveals the presence of at least two molecular outflows in the region, one in an east--west (E--W) direction and the other one in  an almost north-south (N--S) direction. To study the outflows on scales comparable to the HMC, the ALMA correlator was set up to cover the SiO emission, a well-known shock tracer.  As seen in Figs.~\ref{fig-spectra} and \ref{fig-spectra-NE}, the SiO\,(5--4) emission has been clearly detected towards both the Main and the NE core. The spectral profiles of SiO at different positions in the region clearly show broad wings typical of shocked material (Fig.~\ref{fig-sio}).  As seen in
this figure, the SiO profile at the dust continuum emission peak is also affected by absorption. The SiO emission averaged on the blue-shifted and red-shifted wings reveals the presence of several bipolar outflows in the region. In Fig.~\ref{fig-sio-dir}, we have indicated the axes of the possible outflows associated with the core. The directions have been roughly estimated by joining the blue-shifted peaks with what we believe to be the corresponding red-shifted peaks. Another possible interpretation of the blue and red-shifted emission in Fig.~\ref{fig-sio-dir} is that the SiO line is tracing a single, wide-angle bipolar outflow
oriented NE--SW. In this scenario, the two lobes would be heavily resolved out by the interferometer and we detect only their borders. However, we tend to exclude this possibility, because such a prominent outflow should be clearly seen in single-dish maps of the region, in contrast
with the results obtained by Cesaroni et al. (2011). These authors imaged the $^{12}$CO and $^{13}$CO\,(2--1) line emission towards the G31 region with
the IRAM 30-m telescope, and find no obvious evidence of a bipolar outflow on scales $>$5$''$ ($>$0.2~pc). 

The most clear bipolar outflow is that oriented in the E--W direction, whose geometrical center is displaced $\sim$$0\farcs6$ ($\sim$4700\,au) to the south of the dust continuum emission peak and  free-free sources (see Fig.~\ref{fig-sio-dir}). Note that the eastern red-shifted lobe of this outflow could be contaminated by red-shifted emission from another outflow in the region oriented NE--SW  (see below). This could explain why the red lobe of the E-W outflow appears slightly bent towards south.  The extent of this E--W outflow is $\sim$$8\farcs5$ ($\sim$0.33\,pc). The fact that the blue-shifted and red-shifted emissions overlap in the lobes, especially in the western one, suggests that the outflow lies close to the plane of the sky.  

Another bipolar outflow is elongated in an almost N--S direction, with red-shifted and blue-shifted knots located 7$''$--8$''$ from the center of the core. The extent of this bipolar outflow is $\sim$$15''$ ($\sim$0.58\,pc). This N--S SiO outflow is also visible in the $^{12}$CO\,(2--1) maps of Cesaroni et al.~(\cite{cesa11}). The outflow extends further north in SiO than in CO,  probably due to the lower sensitivity of the SMA CO observations. If one traces a line from the southernmost blue-shifted knot to the northernmost red-shifted knot, the  outflow seems to cross the center of the Main core, which suggests that it could be driven by  one of the free-free sources. In addition, as seen in Fig.~\ref{fig-sio-dir}, this outflow could be associated with the water maser jet observed by Moscadelli et al.~(\cite{mosca13}). The analysis of the water maser dynamics by Moscadelli et al.~(\cite{mosca13}) demonstrated clearly that the water masers trace expansion, i.e., outflow(s) in the region, while the methanol and hydroxyl maser dynamics are still not fully understood. 

Through inspection of the SiO emission,  one finds a possible third outflow in a NE--SW direction (Fig.~\ref{fig-sio-dir}). This outflow is clearly not associated with the Main core and it could be associated with a southern core not detected in our observations, possibly due to limitation in the dynamical range. The 3$\sigma$ level of the dust continuum emission is 4.5\mjy, which is $\sim$53 times lower than the peak emission of the Main core. This third outflow is also visible in the $^{12}$CO channel maps of Cesaroni et al.~(\cite{cesa11}). As seen in the CO channel maps and in the SiO averaged emission map, the red-shifted NE lobe extends to the north up to the eastern red-shifted lobe of the E--W bipolar outflow.

\begin{figure*}
\centerline{\includegraphics[angle=0,width=16cm,angle=0]{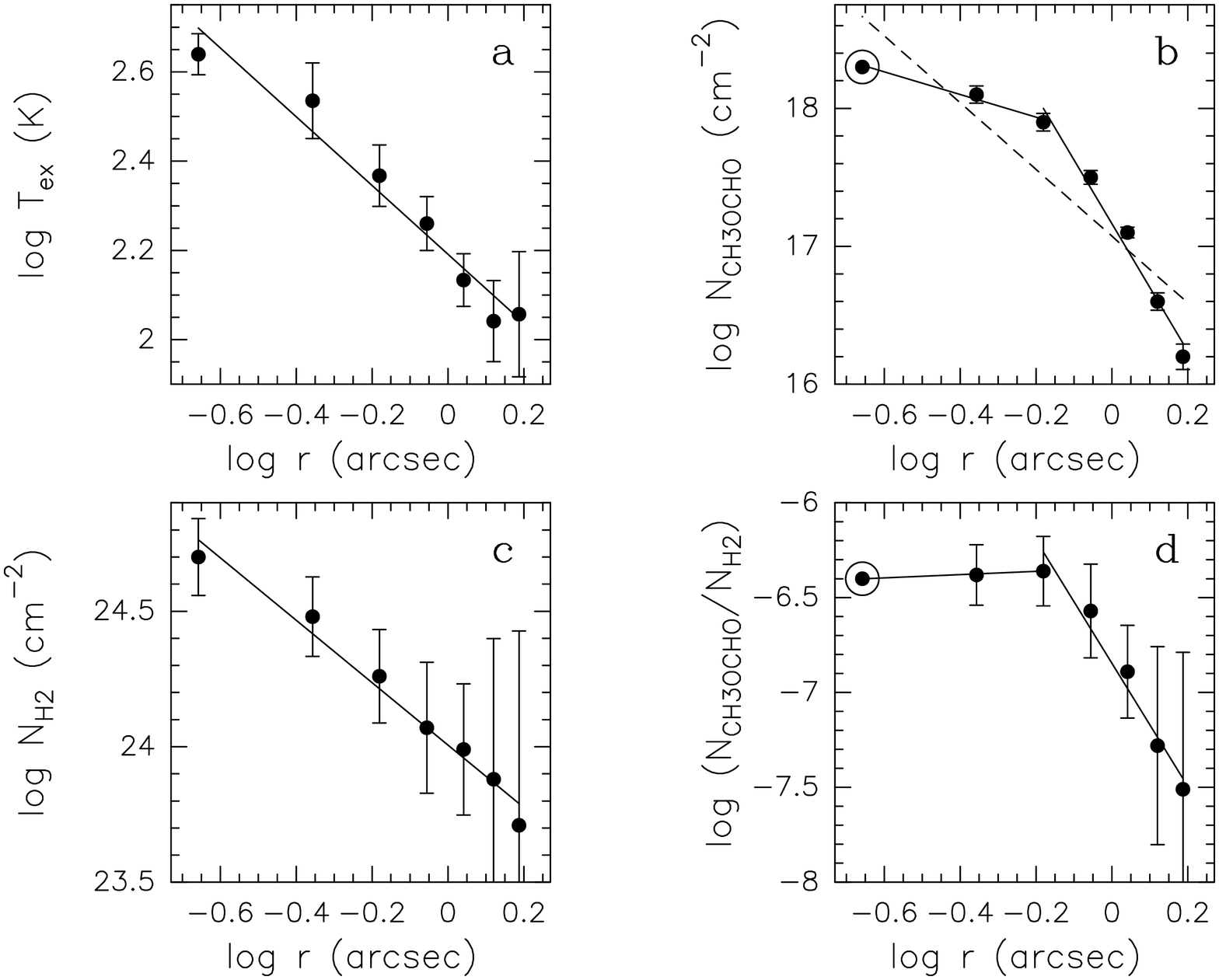}}
\caption{ ({\it a}) Excitation temperature,  ({\it b}) \MF\ column density,   ({\it c}) H$_2$ column density, and  ({\it d}) [$N_{\rm CH_3OCHO}$/$N_{\rm H_2}$] ratio  
as a function of the radius. $T_{\rm ex}$ and $N_{\rm CH_3OCHO}$ have been estimated by fitting the \MF\ $v=0$ and $v_t=1$  emission azimuthally averaged over different rings (see Sect.~\ref{t-gradient}), while $N_{\rm H_2}$ has been estimated from the azimuthally averaged dust emission. The \MF\ column density at a radius of $0\farcs22$ (marked with a circle) has been calculated by extrapolating a linear fit from the $R=0\farcs44$ and $0\farcs66$ rings. The solid lines in 
 {\it a} and {\it c},  and the dashed one in  {\it b} show the least square linear fits to all the points. The solid lines in  {\it b} and {\it d} show the least square linear fits to the inner 3 points and to the outer 5 points in each panel.}
\label{fig-profiles}
\end{figure*}

\begin{figure*}
\centerline{\includegraphics[angle=0,width=14cm,angle=90]{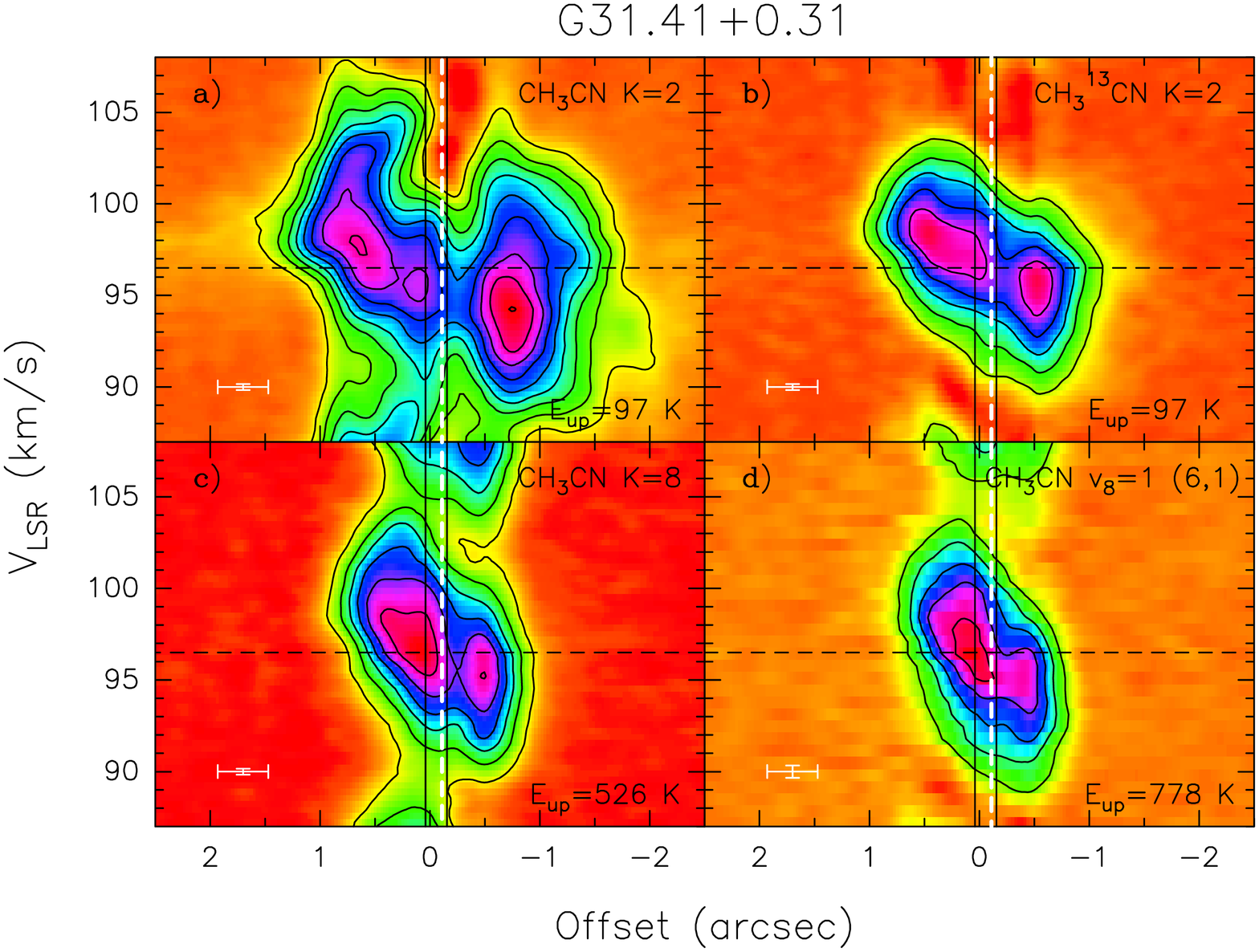}}
\caption{Position-velocity plots along the direction with PA = 68$^\circ$ toward the HMC of a) \MCN\ $K = 2$, b) \MCNII\ $K=2$, c) \MCN\ $K=8$, and d) \MCN\  $K, l=(6, 1)$ $v_8=1$ (12--11). The offsets are measured from the phase center, positive to the northeast. Contour levels range from 20 to 180\,mJy\,beam$^{-1}$ in steps of 20\,mJy\,beam$^{-1}$.  The vertical dashed white line indicates the position of the continuum peak, while the vertical solid black lines show the position of the two compact free-free continuum sources detected by Cesaroni et al.~(\cite{cesa10}). The horizontal dashed line indicates the systemic LSR velocity. The error bars in the lower left-hand corner of the panels indicate the angular and spectral resolution of the data. The upper level energies $E_{\rm up}$ of each transition are indicated in the lower right-hand corner of the  panels. }
\label{fig-pv}
\end{figure*}

\begin{figure}
\centerline{\includegraphics[angle=0,width=9cm,angle=0]{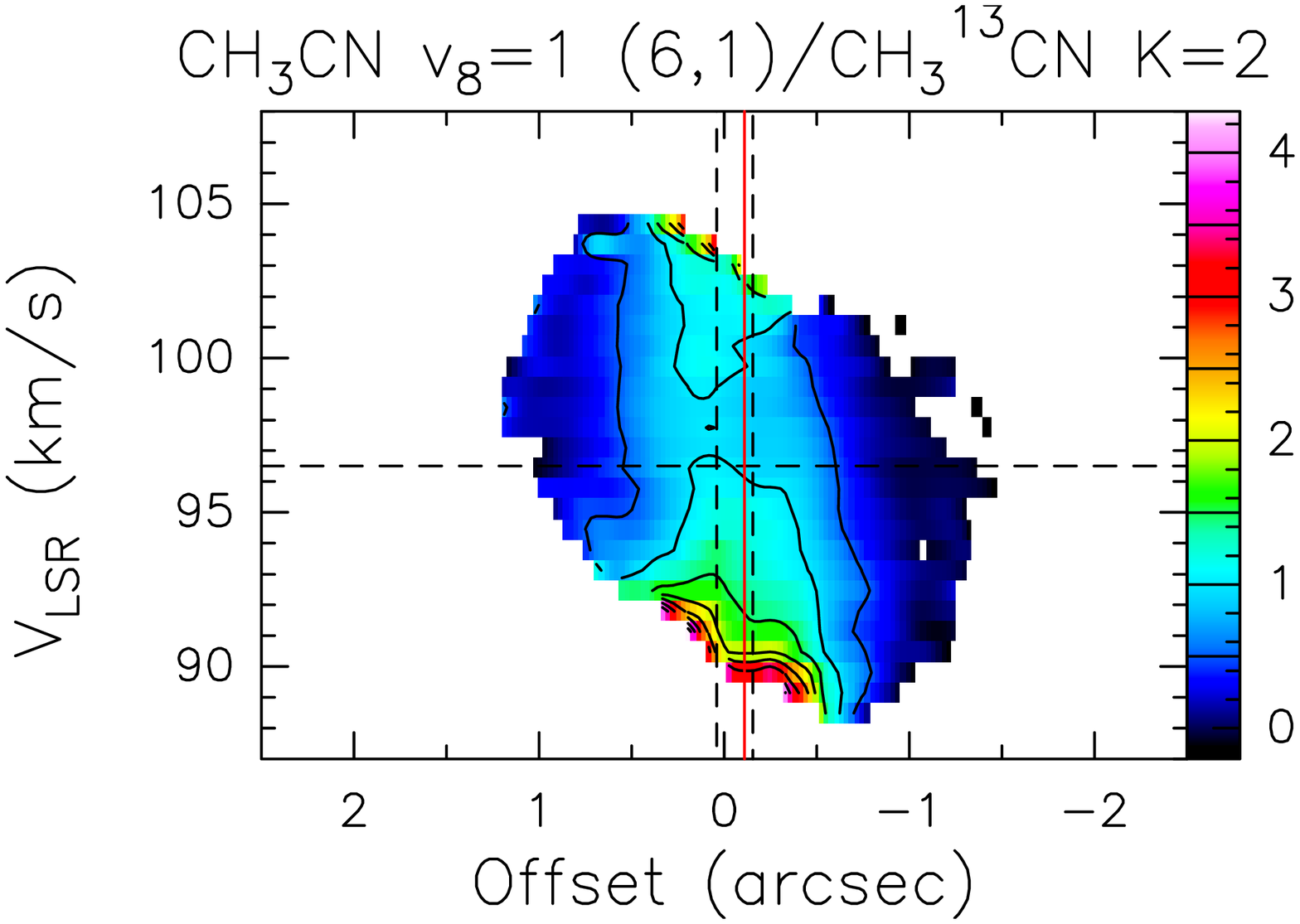}}
\caption{Position-velocity plot along the direction with PA = 68$^\circ$ of the  \MCN\  $K, l=(6, 1)$ $v_8=1$ (12--11) line divided by the same plot of the  \MCNII\ $K=2$ line. The offsets are measured from the phase center, positive to the northeast. Contour levels range from 0.5 to 4 in steps of 0.5.  The vertical solid red line indicates the position of the continuum peak, while the vertical dashed black lines show the positions of the two compact free-free continuum sources detected by Cesaroni et al.~(\cite{cesa10}). The horizontal dashed line indicates the systemic LSR velocity of 96.5\,\kms.}
\label{fig-pv-div}
\end{figure}

\section{Analysis}

\subsection{Fitting the spectra}
\label{fitting}

The high sensitivity of the ALMA observations has allowed us to detect the emission of several transitions of high-density tracers, such as methyl cyanide and methyl formate, with different upper level energies. Given the high density of the core, one could derive the gas temperature and column density by fitting the lines of these molecules with a simple model assuming local thermodynamic equilibrium (LTE) conditions.  Application of this method to all the spectra observed over the core would provide us with maps of excitation temperature, $T_{\rm ex}$, and column density, $N_{\rm tot}$. While software such as {\sc XCLASS} (M\"oller et al.~\cite{moller17}) can automatically perform this fit over large regions corresponding to many spectra, in the case of G31 the situation is more complicated and requires an ad hoc solution. In fact, most line profiles towards the central region of the core are severely affected by deep absorption
(see Fig.~\ref{fig-abs}), which calls for manual fitting to achieve convergence. In order to simplify the problem, we have assumed spherical symmetry
for the G31 core, consistent with its roundish appearance. The core center has been chosen to be the position of the continuum peak and
we have computed mean spectra over circular concentric annuli of width equal to that of the synthesized beam ($\sim$$0\farcs22$), up to a maximum radius of $1\farcs54$, beyond which most of the lines are not detected (see Figs.~\ref{fig-gradients} and \ref{fig-gradients-MF}). 
In this way, by fitting the mean spectra, we have obtained  $T_{\rm ex}$ and $N_{\rm tot}$ as a function of distance from the core center.

In this work, we have used two different programs to fit the lines: {\sc MADCUBA}\footnote{Madrid Data Cube Analysis on ImageJ is a software developed in the
Center of Astrobiology (Madrid, INTA--CSIC) to visualize and analyze
single spectra and datacubes (Mart\'{\i}n et al., in prep.).} (Mart\'{\i}n et al., in prep.; see also Rivilla et al.~\cite{rivilla16}) and  {\sc XCLASS} (M\"oller et al.~\cite{moller17}). Both programs take line blending and optical depth effects into account and in both cases, the emission is described with 5 parameters that can be fixed or left free: the source size $\theta$, the excitation temperature  $T_{\rm ex}$, the column density of the molecule, $N_{\rm tot}$, the line width, $\Delta v$, and the peak velocity of the emission $v$. The continuum emission can be described as a modified blackbody. In general, the outputs of the two codes are consistent. However, in specific cases one of the two algorithms appears to achieve better convergence than the other. In particular, to fit the spectra at the core center, where both emission and absorption components are detected,  {\sc XCLASS} is preferred to {\sc MADCUBA}. The latter package models the core and foreground layers as independent components that do not interact with each other radiatively, whereas in {\sc XCLASS}, the core contribution is simulated as an additional continuum background (M\"oller et al.~\cite{moller17}). This approach resembles our scenario where all the components along the line of sight interact radiatively. On the other hand, when estimating the temperature and column density profiles  across the core (see next section) by fitting a single component in emission, we have used {\sc MADCUBA} because 
the fitting algorithm converges faster than {\sc XCLASS}. It should be noted that the \MF\ emission towards the center, for which the absorption is strongly affecting the profiles of the emission lines, has not been fitted because one needs at least two components. The fitting of the central region is discussed in Sect.~\ref{excitation-absorption}.

\subsection{Temperature and density profiles}
\label{t-gradient}

To estimate the physical parameters across the core, we fitted the ground state rotational transitions of methyl formate ($v=0$) and those in the first torsionally excited state ($v_t=1$). 
The choice of \MF\ instead of \MCN\ is based on the fact that the \MCN\ spectra closer to the central position ($R\leq0\farcs66$) cannot be properly fitted because of the high opacity of the ground state transitions. 
In addition, as already mentioned, we have detected $\sim$40  \MF\ transitions (including both $v=0$ and $v_t=1$) that are either unblended or only slightly blended with other lines,  whereas only few unblended \MCN\ transitions (including $v=0$, $v_8=1$, and isotopologues) could be identified. This allows us to better constrain the fit parameters. 
 The fitting strategy we adopted in {\sc MADCUBA} was to fix the line width to a value that provides a good visual fit, and the source size, that we assumed to be larger than the beam size. For the dust continuum temperature at each annulus, we assumed the value of the azimuthally averaged brightness temperature of the dust emission. Table~\ref{table-mf} shows the best-fit  $T_{\rm ex}$ and $N_{\rm CH_3OCHO}$ values at each radius. A successful fit to the spectra for the innermost radius was not achieved due to the strong absorption, which affects the profiles and prevents the algorithm from converging. The fit of the $R=0\farcs22$ ring gives an unrealistically high excitation temperature of $>1000$\,K if both $T_{\rm ex}$ and $N_{\rm CH_3OCHO}$ are left as free parameters. Therefore, we fixed the column density to $2\times10^{18}$\,cm$^{-2}$, the value obtained from linear extrapolation of the column densities at the two adjacent inner rings.  

\begin{table}
\caption[] {Best fit model parameters for the \MF\ $v=0$ and $v_t=1$ transitions for the emission azimuthally averaged at different rings.}
\label{table-mf}
\begin{tabular}{cccc}
\hline
\multicolumn{1}{c}{$R$}
&\multicolumn{1}{c}{$R$}
&\multicolumn{1}{c}{$T_{\rm ex}$}
&\multicolumn{1}{c}{$N_{\rm CH_3OCHO}$}
\\
\multicolumn{1}{c}{(arcsec)}
&\multicolumn{1}{c}{(au)} 
&\multicolumn{1}{c}{(K)}
&\multicolumn{1}{c}{(10$^{17}$ cm$^{-2}$)}
\\
\hline
0.22 &1740 &436$\pm$20 &20$^{\rm a}$   \\
0.44 &3480 &343$\pm$29 &12.6$\pm$0.8   \\
0.66 &5200 &233$\pm$16 &7.9$\pm$0.5   \\
0.88 &7000 &182$\pm$11 &4.0$\pm$0.2   \\
1.10 &8700 &136$\pm$8 &1.45$\pm$0.05  \\
1.32 &10430  &110$\pm$10 &0.48$\pm$0.03   \\
1.54 &12170  &114$\pm$16 &0.16$\pm$0.014  \\
\hline
\end{tabular}

$^a$ This value has been fixed to the value obtained by extrapolating a linear fit from the $R=0\farcs44$ and $0\farcs66$ rings. If both $T_{\rm ex}$ and $N_{\rm CH_3OCHO}$ are left as free parameters, the algorithm gives an unrealistic high excitation temperature of $>1000$\,K for the $R=0\farcs22$ ring. \\

\end{table}

Figure~\ref{fig-profiles} shows the temperature and column density, estimated by fitting the \MF\ $v=0$ and $v_t=1$ emission, as a function of radius, $R$, and the least-square linear fits. The best fit to the $\log (T_{\rm ex}$) vs. $\log (R)$ relation is:
\begin{equation}
\log (T_{\rm ex}/{\rm K}) = (2.19\pm0.02) - (0.77\pm0.07)\,\log (R/{\rm arcsec}) 
\label{tex}
\end{equation}

The fit implies that $T_{\rm ex}$ is proportional to $R^{-0.77}$. This temperature profile is much steeper than the $R^{-0.4}$ profile found at the outer parts of envelopes surrounding massive YSOs by van der Tak et al.~(\cite{vandertak00}), but is consistent with the profiles of hot molecular cores modeled by Nomura \& Millar (\cite{nomura04}) and Osorio et al.~(\cite{osorio09}). Alternatively, the $T_{\rm ex}$ profile would also be consistent with those estimated for passive, geometrically flat, irradiated dust disks (Adams \& Shu~\cite{adams86}) or steadily accreting, thin disks (Shakura \& Sunyaev~\cite{shakura73}),  which have power-law exponents of $-0.75$. 

As shown in Fig.~\ref{fig-profiles}b, the CH$_3$OCHO column density profile is best fitted by a broken  power law. The best fit to the  
$\log (N_{\rm CH_3OCHO}$) vs. $\log (R)$ relation is:

\begin{multline}
\log (N_{\rm CH_3OCHO}/{\rm cm^{-2}}) = 
\\
\left\lbrace
\begin{array}{r@{}l}
 (17.17\pm0.05) - (4.63\pm0.41)\,\log (R/{\rm arcsec})     \hspace{.4cm}    R > 0.66'' 
 \\
 \\
  (17.77\pm0.06) - (0.82\pm0.12)\,\log (R/{\rm arcsec})   \hspace{.4cm}    R \leq 0.66''
 \end{array}   
\right. 
\end{multline}

In the inner part of the core, the column density is proportional to $R^{-0.8}$, much flatter than the outer part of the core, which shows a very steep profile  $R^{-4.6}$. The fact that the $N_{\rm CH_3OCHO}$ profile flattens towards the center of the core is likely an effect of absorption, which is affecting the \MF\ line profiles in the inner rings ($R<5000$\,au). 
To study whether the drastic drop of \MF\ column density is due to a decrease of the \MF\ abundance or is produced by a drop of the gas density towards the center of the core, we estimated the H$_2$ column density from the dust continuum emission azimuthally averaged over the same concentric rings as the \MF\ emission. We assumed that the dust temperature is equal to the excitation temperature estimated for \MF\ at each ring (see Table~\ref{table-mf}). Figure~\ref{fig-profiles}$c$ shows $N_{\rm H_2}$ as a function of $R$. As seen in the figure, $\log (N_{\rm H_2}$) follows a linear correlation with $\log (R)$. The best fit is:
\begin{equation}
\log (N_{\rm H_2}/{\rm cm^{-2}}) = (24.01\pm0.02) - (1.15\pm0.08)\,\log (R/{\rm arcsec}) 
\label{nh2}
\end{equation}

The H$_2$ column density seems to be proportional to $R^{-1.2}$, and the profile does not  show an abrupt change of slope at $R>5000$\,au like that observed in  \MF. 
The dust emission at inner radii is optically thick, as indicated by the high brightness temperature of the continuum peak ($\sim$132\,K) being comparable to the gas (and dust) temperature ($\gtrsim$150\,K). Therefore, the H$_2$ column density values should be taken as lower limits. However, the slope of the power-law distribution is very steep ($-1.15$), and therefore, the correction due to the large continuum opacity should likely not affect the estimated column density dramatically. Otherwise, the volume density profile would be extremely steep ($\propto r^{-n}$ with $n>2.2$),  difficult to justify in (roughly) spherical geometry either for a static envelope or an accretion flow (e.g., van der Tak et al.~\cite{vandertak00}).  This suggests that the drop seen in \MF\ column density is due to a decrease of its abundance, as shown by the [$N_{\rm CH_3OCHO}$/$N_{\rm H_2}$] ratio in Fig.~\ref{fig-profiles}$d$. While for $R<5000$\,au the \MF\ abundance is almost constant with a value of $\sim4\times10^{-7}$, for  $R>5000$\,au the abundance is proportional to $R^{-3.2}$ and drops almost an order of magnitude at the outer rings.  An even more drastic decrease of the abundance at the outer part of the core is observed for \MCN. Unfortunately, for this species, it has not been possible to fit the emission for the rings with $R<7000$\,au, so it is not possible to investigate whether the abundance is constant  at the center of the core. The best least-square linear fits to the \MF\ abundance are:

\begin{multline}
\log [N_{\rm CH_3OCHO}/N_{\rm H_2}] = 
\\
\left\lbrace
\begin{array}{r@{}l}
 -(6.85\pm0.05) - (3.23\pm0.36)\,\log (R/{\rm arcsec})     \hspace{.4cm}    R > 0.66'' 
 \\
 \\
 -(6.35\pm0.01) + (0.08\pm0.01)\,\log (R/{\rm arcsec})   \hspace{.4cm}    R \leq 0.66''
 \end{array}   
\right. 
\end{multline}

The rich chemistry of hot molecular cores, in particular of complex molecules such as \MF, is the result of the evaporation of dust grain mantles. One possibility to explain the drastic drop of the \MF\  abundance could be that the temperature is not high enough to desorb \MF\ from the dust grains. However, Burke et al.~(\cite{burke15}) have conducted laboratory experiments and estimated that \MF\ desorbs at a temperature of about 70\,K, which is much lower than the excitation temperatures estimated in the outer rings of G31 ($>110$\,K). Therefore, the only plausible explanation for the abundance drop observed in the G31 core is that \MF\  is concentrated in the regions closer to the central protostar.

\subsection{Mass estimates}

Previous mass estimates of \g31 range from $\sim$500 to 1700\,$M_\odot$ (Beltr\'an et al.~\cite{beltran04}; Girart et al.~\cite{girart09}; Cesaroni et al.~\cite{cesa11}).  These masses have been derived from the dust continuum emission, assuming a single value of the temperature. The large discrepancy is mainly due to the different dust absorption coefficient used by the different authors. Because we have a better knowledge of the physical properties of the core, having  derived the temperature and density distribution of the core (see previous section), we can improve the mass estimate of \g31. To do this, we estimated the mass of the core, $M_c$, assuming that the dust emission is optically thin and in the Rayleigh-Jeans regime (in fact, $\left[\frac{\nu}{\rm GHz}\right]\ll 20\,\left[\frac{T_{\rm ex}}{{\rm K}}\right]$), and that the density and temperature of the core follow power-law distributions, $\rho\propto r^{-n}$ and $T\propto r^{-q}$. By integrating the density distribution over the volume, one finds that 
$M_c = \frac{4\pi}{3-n}\rho_c\,r_c^3$, where $\rho_c$ is the density at the radius of the core, $r_c$. Following Eqs. (1) and (2) of Beltr\'an et al.~(\cite{beltran02}), the density at the radius of the core can be written as
\begin{equation}
\rho_c = \frac{1}{4\pi\sqrt\pi}\frac{\Gamma[(n+q)/2]}{\Gamma[(n+q-1)/2]}\frac{3-n-q}{r_c^3}\frac{c^2}{k\,\nu^2\,T_c\,\kappa_\nu}\,d^2\,S_\nu,
\label{rho}
\end{equation}
where $\Gamma(x)$ is the gamma function, $\kappa$ is the Boltzmann constant, $\nu$ is the frequency of the observations, $T_c$ is the temperature at the radius of the core, $\kappa_\nu$ is the dust absorption coefficient per unit mass, $d$ is the distance of the source, and $S_\nu$ is the integrated flux density over the whole core. Using this equation, $M_c$ can be written as
\begin{equation}
M_c =  \frac{1}{\sqrt\pi}\frac{\Gamma[(n+q)/2]}{\Gamma[(n+q-1)/2]}\frac{3-n-q}{3-n}\frac{c^2}{k\,\nu^2\,T_c\,\kappa_\nu}\,d^2\,S_\nu.
\end{equation}

 We estimated $\rho_c$ and $M_c$ for the G31 Main core, for a radius $r_c$ of 8500\,au ($\sim$$1\farcs076$), which is the deconvolved radius of the dusty core at the 5$\sigma$ emission level. For this purpose, we adopted $q$=0.77 and $T_c$=146\,K (see Eq.~[\ref{tex}]), $\nu$=217\,GHz, $k_\nu$=0.008\,cm$^{2}$\,g$^{-1}$ (Ossenkopf \& Henning~\cite{ossenkopf94} and a gas-to-dust mass ratio of 100), and $S_\nu=3.10$\,Jy (see Table~\ref{table-cont}).  The H$_2$ column density profile (Eq.~[\ref{nh2}]) implies a density power-law exponent $n\sim2.2$. On the other hand, the red-shifted absorption observed in several tracers suggests that the core might be undergoing free-fall collapse (see Sect.~\ref{high-tracers} and \ref{red-shifted}) with $n\sim1.5$. Therefore, we assumed an intermediate value of $n=2$.
This leads to $\rho_c$=$9\times10^{-18}$\,g\,cm$^{-3}$, corresponding  to a volume density of $\sim$$5.4\times10^6$\,cm$^{-3}$, and $M_c$$\simeq120\,M_\odot$. 
Because the mass has been estimated assuming that the dust emission is optically thin, the value derived has to be taken as a lower limit in case this condition is not satisfied.

\section{Discussion}

\subsection{The NE--SW velocity gradient}

Figure~\ref{fig-gradients} shows the 217\,GHz continuum emission overlaid on the integrated intensity (moment 0), line velocity (moment 1), and velocity dispersion (moment 2) maps of different transitions of \MCN\ (ground state and vibrationally excited) and isotopologues.  These transitions have been selected to trace emission with upper level energies ranging from 97 to 778 K. The emission of methyl cyanide shows a clear velocity gradient along the NE--SW direction regardless of the upper level energy of the transition.  The position angle of the velocity gradient is $\sim$68$^\circ$ and is roughly consistent with that of the HMC dust continuum emission,  PA$\sim$63$^\circ$ (see Sect.~\ref{cont-em}).  The same velocity gradient is observed in \MF\ (Fig.~\ref{fig-gradients-MF}). This velocity gradient, already observed in \MCN\ by Beltr\'an et al.~(\cite{beltran04}) and Cesaroni et al.~(\cite{cesa11}) and in CH$_3$OH  by Girart et al.~(\cite{girart09}) has been interpreted as rotation of the core. The velocity gradient is $\sim$50\,\kms\,pc$^{-1}$ for \MCN\ and \MCNII\ $K$=2 and increases for the higher  energy transitions, being $\sim$87\,\kms\,pc$^{-1}$ for $K$=8 and $\sim$107\,\kms\,pc$^{-1}$ for $K, l=(6, 1)$ $v_8=1$.

\subsubsection{Rotational spin-up}

The velocity gradient is also evident in Fig.~\ref{fig-pv}, which shows the position-velocity (PV) plots along PA = 68$^\circ$ for the same transitions as in Fig.~\ref{fig-gradients}.  The \Vlsr\ appears to increase linearly with distance from the core center, consistent with solid-body rotation. Figure~\ref{fig-pv} also shows the \MCN\ $K$=2 line absorption at red-shifted velocities. By comparing the different PV plots, one sees that the rotation seems to speed up with increasing energy of the transition. The effect is especially remarkable if one compares the PV plot of the vibrationally excited \MCN\,(12--11)  $K, l=(6, 1)$ $v_8=1$ ($E_{\rm up}$= 778\,K) line with the PV plot of the ground-state \MCNII\,(12--11) $K$=2 ($E_{\rm up}$= 97\,K)  line. The former looks steeper than the latter.  The \MCN\ $K$=2 transition is not considered because it is too affected by red-shifted absorption. The same steepness is also visible in the PV plot of the $K$=8 transition of \MCN\,(12--11) ($E_{\rm up}$= 526\,K) line. The PV  plot of the vibrationally excited line shows emission that extends up to $\sim\pm0\farcs65$ from the central position and has velocities ranging from 89 to 104\,\kms, while the emission in the \MCNII\ $K=2$ transition extends up to $\sim\pm1\farcs1$ from the center and the velocities range from 90 to $\sim$103.5\,\kms. This corresponds to slopes of $\sim$300\,\kms\,pc$^{-1}$ and $\sim$160\,\kms\,pc$^{-1}$, respectively. This change in slope of the PV plots of the low and high energy transitions is emphasized in Fig.~\ref{fig-pv-div}, which shows the PV plot along the direction with PA = 68$^\circ$ of the ratio between the \MCN\  $K, l=(6, 1)$ $v_8=1$ (12--11) and the  \MCNII\ $K=2$ line. The ratio is largest at small offsets and smallest at large radii. If the gradient is interpreted as  rotation, this PV plot suggests that the rotation increases with increasing energy of the transition, that is, towards the core center. This is expected for conservation of angular momentum in a rotating and infalling structure. This result seems to be inconsistent with the solid-body rotation suggested by the PV in each line (see above).  A possible explanation is that each line is tracing only a relatively narrow annulus of material where the gas temperature maximizes the emission from that line due to the fact that such temperature is comparable to the line upper level energy. This would explain why the projection of the rotation velocity along the line of sight increases linearly with distance form the star, mimicking solid-body rotation. At the same time, higher energy lines trace smaller annuli, closer to the star and thus rotating faster, which explains the increasing steepness of the PV plot with increasing line energy. 

A scenario in which rotation spins up close to the center of the core is  different from that depicted by Girart et al.~(\cite{girart09}), who find evidence of decreasing rotation velocities towards the center of G31 and suggest that such a spin-down could be produced by magnetic braking. Girart et al.~(\cite{girart09}) estimated the rotation velocities at the position where the dust major axis intersects the 50\% contour level in the PV plots of methanol transitions with different upper level energies. On the other hand,  from Fig.~3 of Girart et al.~(\cite{girart09}), one can see that the PV plot of CH$_3$OH $v_t=2$ ($7_5-6_5$), which has $E_{\rm up}>$ 900\,K is clearly steeper than that of CH$_3$OH $v_t=2$ ($3_3-4_2$)E, with a significantly lower energy $E_{\rm up}\sim$60\,K, in agreement with our findings. This suggests that the discrepancy between our interpretation and that of Girart et al.\ is not the result of the choice of the species used to study the rotation velocity but of the method used to analyze the data. Given the uncertainty in estimating the size and corresponding velocity from the 50\% contour level in a PV plot, we believe that the rotation curve obtained by Girart et al.\ is not to be taken as strong evidence against our interpretation of  increasing rotation speed towards the center of the core. 

It is also worth analyzing the velocity dispersion in the HMC. The bottom panels of Fig.~\ref{fig-gradients} show the maps of the 2nd moment of the methyl cyanide lines. For the transitions less affected by red-shifted absorption, namely those with higher energy, the velocity dispersion peaks towards the position of the dust continuum emission peak, as expected for rotation spin up.  For  \MCN\ and  \MCNII\ $K=2$, clearly affected by absorption, the velocity dispersion increases in the directions associated with the molecular outflows in the region.

\subsubsection{Interaction with the molecular outflow}

Figure~\ref{fig-sio-dir} shows the presence of at least three outflows in the region. It has been hypothesized that the NE--SW velocity gradient seen in high-density tracers could indeed be produced by one of these outflows (Gibb et al.~\cite{gibb04}; Araya et al.~\cite{araya08}). The question was investigated  by Cesaroni et al.~(\cite{cesa11}), who concluded that although the rotation scenario appeared more plausible, the outflow interpretation could not be totally discarded. With the increased angular resolution of  our ALMA observations, we can now reconsider the possible relationship between the outflows in the region and the velocity gradient. 

\begin{figure}
\centerline{\includegraphics[angle=0,width=9cm,angle=0]{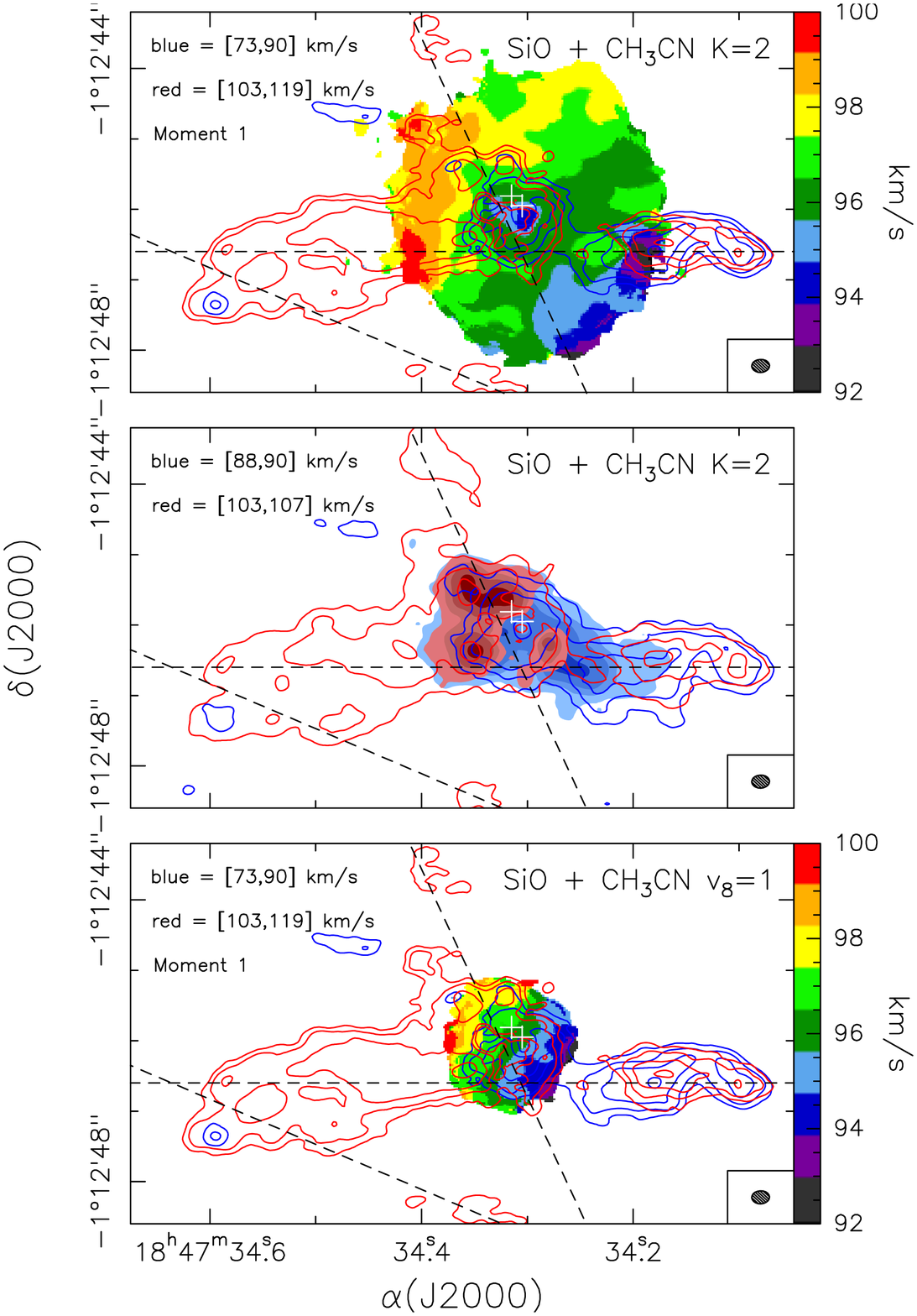}}
\caption{{\it Top panel}: overlay of the line velocity (moment 1) map of CH$_3$CN $K=2$ (12--11) ({\it colors}) on the SiO\,(5--4)  blue-shifted ({\it blue contours}) and red-shifted ({\it red contours}) averaged emission. The CH$_3$CN $K=2$ line velocity map has been computed over the velocity range (92, 100)\,\kms. The velocity intervals over which the SiO emission has been averaged are indicated in the upper left-hand corner.  Contour levels are the same as in Fig.~\ref{fig-sio}. {\it Middle panel}: overlay of the CH$_3$CN $K=2$ (12--11)  and SiO\,(5--4) blue-shifted and red-shifted emission averaged over the same velocity intervals, which are indicated in the upper left-hand corner. Contour and grayscale levels are 6, 18, 36, and 72\,mJy\,beam$^{-1}$. {\it Bottom panel}: same as the {\it top panel} but for CH$_3$CN $K, l = (6, -1)$ $v_8$=1  (12--11). The synthesized beam is shown in the lower right-hand corner.  The white crosses indicate the positions of the two compact free-free continuum sources detected by Cesaroni et al.~(\cite{cesa10}). The black dashed lines indicate the direction of the three possible outflows.}
\label{fig-sio-comp}
\end{figure}

The bipolar outflow that could be interacting with the core material and producing the NE--SW gradient would be what we have called outflow E--W (Fig.~\ref{fig-sio-dir}). This outflow has been traced in $^{12}$CO and $^{13}$CO by Cesaroni et al.~(\cite{cesa11}) and, as seen in their lower angular resolution maps, it appears to have its geometrical center close to the center of the core. However, as seen in Fig.~\ref{fig-sio-comp}, when observed at higher angular resolution, the outflow appears clearly displaced from the center of the core and not associated with any of the embedded free-free  sources nor the dust continuum peak, whereas the velocity gradient seems to be centered at the dust continuum peak.  Furthermore, the direction of the outflow is significantly different from that of the velocity gradient: whereas the former  
is oriented in an E--W direction, the latter, when traced in the higher energy \MCN\ $v_8=1$ transitions, is oriented NE--SW  (see Fig.~\ref{fig-sio-comp}). This indicates that the motion of the innermost material close to the center (traced by the highest energy transitions) is not consistent with that of the bipolar outflow.  Figure~\ref{fig-sio-comp} shows the superposition of the velocity gradient traced in \MCN\ $K=2$ (top panel) and the SiO molecular outflow. As one can see, the $K=2$ transition traces much more extended, and possibly less dense, material than the  $v_8=1$  transition. The line velocity (moment 1)  map of \MCN\ shows a  red-shifted peak at $\sim$100\,\kms\ and a blue-shifted peak at $\sim$92\,\kms\ coincident with the lobes of the bipolar outflow and aligned in an E--W direction, which suggests that the bipolar outflow could be indeed interacting with the outer regions of the core. If the \MCN\ $K=2$ emitting gas is entrained by the molecular outflow, then one would expect this gas to be tracing the high-velocity material in a similar way to SiO. To check if this is the case,  we averaged both the high-velocity SiO and \MCN\ $K=2$ emission in the same velocity intervals: (88, 90)\,\kms\ for the blue-shifted emission and (103, 107)\,\kms\ for the red-shifted emission. As seen in the middle panel of Fig.~\ref{fig-sio-comp}, the \MCN\ $K=2$ high-velocity emission indeed traces the inner region of the molecular outflow, especially at blue-shifted velocities. We have also averaged the  \MCN\ $v_8=1$ in the same high-velocity intervals, but the emission is very weak and mostly concentrated towards the center or the core, with no coincidence with the outflow lobes. This suggests that the velocity gradient traced in the lowest energy  transitions is in part due to expansion (at least for the high velocities). On the other hand, the velocity gradient seen in the highest energy lines, coming from the innermost, hottest layers of the core, does not appear to be affected by the outflow and could be tracing rotation.

Following Cesaroni et al.~(\cite{cesa11}), we estimated the outflow parameters from the SiO and \MCN\ $v_8=1$ lines to check whether the velocity gradient traced in  \MCN\ $v_8=1$ could be produced by the expansion of the outflow. In our calculations, we assumed a range of temperatures of 20--50\,K for SiO and 100--250\,K for vibrationally excited  \MCN.  The lower limit for the SiO temperature is set by the peak brightness temperature (see Fig.~\ref{fig-sio}), while for \MCN\ $v_8=1$ the lower limit is the temperature adopted by Cesaroni et al.~(\cite{cesa11}) to calculate the parameters from \MCN\ $K=4$ and \MCNII\ $K=2$. The  abundance of \MCN\  relative to H$_2$ was assumed to be equal to $10^{-8}$  (see e.g.\ van Dishoeck et al.~\cite{vandishoeck93}), while for  SiO we used a range of  $10^{-8}$--$10^{-7}$, following Codella et al.~(\cite{codella13}). Table~\ref{table-outflow} gives the mass of the outflow, $M$, the momentum, $P$, the energy, $E$, and the corresponding rates obtained by dividing the previous quantities by the dynamical time scale of the outflow, $t_{\rm dyn}\simeq6\times10^3$\,yr, which has been calculated as $R_{\rm lobe}/V_{\rm max}$, where $R_{\rm lobe}$ is the size ($\sim$0.15\,pc) of the SiO lobes and $V_{\rm max}$ is the maximum velocity ($\sim$23\,\kms) of the SiO outflow emission with respect to the systemic LSR velocity of 96.5\,\kms.

\begin{table}
\caption[] {E--W outflow parameters calculated from SiO and \MCN\ $v_8=1$.}
\label{table-outflow}
\tabcolsep 3pt 
\begin{tabular}{lcc}
\hline
&\multicolumn{1}{c}{SiO (5--4)$^a$}
&\multicolumn{1}{c}{\MCN$^b$ (12--11)} \\
\multicolumn{1}{c}{Parameters}
&&
\multicolumn{1}{c}{ $v_8=1$ $K, l = (6, 1)$}
\\
\hline
$M (M_\odot)$                                   & 0.36--3.8 &274--7000   \\
$P (M_\odot\,\kms)$                          &4.5--47  &(1--27)$\times10^3$  \\
$E (L_\odot\,$yr)                               &( 0.50--5.3)$\times10^4$  & (0.4--10)$\times10^6$   \\
$\dot M (M_\odot\,$yr$^{-1})$           &(0.60--6.3)$\times10^{-4}$  &  0.05--1.2 \\
$\dot P (M_\odot\,\kms$\,yr$^{-1})$  &(0.75--7.8)$\times10^{-3}$ & 0.17--4.5   \\
$\dot E (L_\odot)$                             & 0.83--8.8   &67--1667  \\
\hline
\end{tabular}
\\
$^a$ Parameters estimated assuming a range of gas temperature of 20--50\,K and a range of abundance relative to H$_2$ of $10^{-8}$--$10^{-7}$. The blue-shifted velocity interval is $\sim$(73--90)\,\kms, and the red-shifted one  $\sim$(103--119)\,\kms.\\
$^b$ Parameters estimated assuming a range of gas temperature of 100--250\,K, and an abundance relative to H$_2$ of $10^{-8}$. The blue-shifted velocity interval is  $\sim$(88--95)\,\kms, and the red-shifted one  $\sim$(98--105)\,\kms.
\end{table}

The outflow parameters estimated from SiO for a gas temperature of 50\,K are similar to those calculated by Cesaroni et al.~(\cite{cesa11}) from $^{12}$CO for  a gas temperature of 100\,K, while those estimated from \MCN\ $v_8=1$ $K, l = (6, 1)$ for a gas temperature of 250\,K are similar to those obtained by Cesaroni et al.~(\cite{cesa11}) from \MCNII\  $K=2$ for  a gas temperature of 100\,K.  As seen in Table~\ref{table-outflow}, the outflow parameters estimated from vibrationally excited \MCN\ are orders of magnitude higher than those estimated from SiO, a typical outflow tracer, and therefore, not realistic. Such high values could only be consistent with those estimated, by means of single-dish observations, for pc-scale molecular outflows powered by O-type protostars (e.g., L\'opez-Sepulcre et al.~\cite{sepulcre09}). In addition, note that our ALMA observations can miss extended emission, and therefore, the values estimated by us from \MCN\ $v_8=1$ are likely lower limits.

 Our results suggest that \MCN\ and SiO are tracing different gas with different motions and confirm the results of Cesaroni et al.~(\cite{cesa11}), who concluded that it is very unlikely that the velocity gradient traced in \MCN\ and isotopologues is due to the expansion of the molecular gas.  We conclude that this velocity gradient is produced by rotation of the core.

\subsection{Red-shifted absorption}
\label{red-shifted}

The most characteristic feature of the line emission towards the Main core in \g31 is the inverse P-Cygni profile observed, although with different degrees  of absorption, in basically all the species and at all  energies. This red-shifted absorption was not detected in previous observations of the same lines, in particular \MCN\ and \MCNII\ (12--11) with the SMA and IRAM PdBI interferometers, due to the limited angular resolution. The detection of red-shifted absorption towards the core center (see Fig.~\ref{fig-abs}) strongly supports the existence of infall in the core. 

Following Beltr\'an et al.~(\cite{beltran06}), we estimated the infall rate inside a solid angle $\Omega$ for a radius of $1\farcs54$ ($\sim$$1.2\times10^4$\,au), which is that of the \MCN\ $K=2$ integrated emission (see top left panel of Fig.~\ref{fig-gradients}), and an infall velocity, assumed to be equal to the difference between the velocity of the absorption feature and the systemic LSR velocity, of 5.1\,\kms. The choice of \MCN\ $K=2$ is based on the fact that red-shifted absorption is very pronounced in this transition. The infall  accretion rate  is $\dot M_{\rm inf}=\Omega/(4\,\pi)\,3\times10^{-2}$\,$M_\odot\,$yr$^{-1}$.  Such a high value is a factor 10 higher than that estimated from modeling (Osorio et al.~\cite{osorio09}) for an infalling envelope of 2.3$\times10^4$\,au, but is consistent with the value estimated by Girart et al.~(\cite{girart09}) at 345\,GHz. Similar high infall rates have been estimated for other O-type (proto)stars (see Beltr\'an \& de Wit~\cite{beltran16}). 

\subsubsection{Accelerating infall}

To study whether the absorption changes with the energy of the line, we measured the velocity of the absorption feature in transitions spanning a broad range of energies at the position of the dust continuum peak. Table~\ref{table-vred}  gives  the different species and transitions, the upper level energy of the line, and the difference between 
the velocity of the red-shifted absorption dip and  the systemic LSR velocity (96.5\,\kms). The red-shifted velocity has only been calculated for those lines that are not heavily blended with other species, because the blending affects the estimate. For a few transitions that are slightly blended, we estimated a lower limit of the velocity.  In Fig.~\ref{fig-vred}, we plot the red-shifted velocity as a function of the upper level energy of the \MCN\ (ground state and vibrationally excited) lines,  \MCNII, \MCNI,  and H$_2$CO. The red-shifted velocity is lower for the H$_2$CO transitions, which have the lowest energies.  The H$_2$CO lines are heavily affected by the absorption as suggested by both the extension and the depth of the absorption (see Fig.~\ref{spectra-h2co}). In particular, H$_2$CO\,(3$_{0, 3} - 2_{0, 2}$), the lowest energy transition, shows no emission at the channel with the deepest absorption, and the brightness temperature $T_{\rm B}$ of the absorption is $-$134\,K, comparable in absolute value to the brightness temperature of the continuum emission peak. Focusing on \MCN\ and isotopologues, Table~\ref{table-vred} and Fig.~\ref{fig-vred} show that the velocity of the absorption feature increases with the transition upper level energy, which suggests that the infall is accelerating towards the core center (where the temperature is higher), consistent with free-fall collapse. It should be noted that the red-shifted velocities of the \MCNI\,(13--12) lines are slightly higher than those of \MCN\ and \MCNII\,(12--11) with same $K$ transition and similar energy. This suggests that the conditions to excite the  (13--12) transitions occur deeper in the core and closer to the center. In any case, the increase of velocity with energy is found also if one considers only the  (13--12) lines.

\subsubsection{Excitation conditions at the core center}
\label{excitation-absorption}

We used the maps with the continuum not subtracted to study the excitation conditions of the gas at the core center, namely, in the inner $0\farcs22$ of the core, where the absorption is strongest. As mentioned before, the species most affected by  red-shifted absorption are \MCN\ and isotopologues, and H$_2$CO. Because \MCN\ has more transitions that can be fitted simultaneously than H$_2$CO, and the spectral resolution of the observations is higher (0.33 to 0.66\,\kms\ for methyl cyanide and isotopologues, 2.7\,\kms\ for formaldehyde), we modeled the absorption towards the center of the G31 Main core in  \MCN\ and its isotopologues.  To estimate the physical parameters of the species, we modeled the spectrum at the central pixel with three different components: i) a background layer with uniform intensity to describe the dust continuum core, ii) a layer to describe the core emission, and iii) a foreground layer located in front of the core component to describe the absorption. For the modeling, we used {\sc XCLASS} as explained in Sect.~\ref{fitting}.
The core and the foreground components are described with 5 parameters each: the source size $\theta$, the excitation temperature  $T_{\rm ex}$, the column density of the molecule, $N_{\rm tot}$, the line width, $\Delta v$, and the offset velocity relative to the \Vlsr, $v_{\rm off}$.

\begin{figure}
\centerline{\includegraphics[angle=0,width=9cm,angle=0]{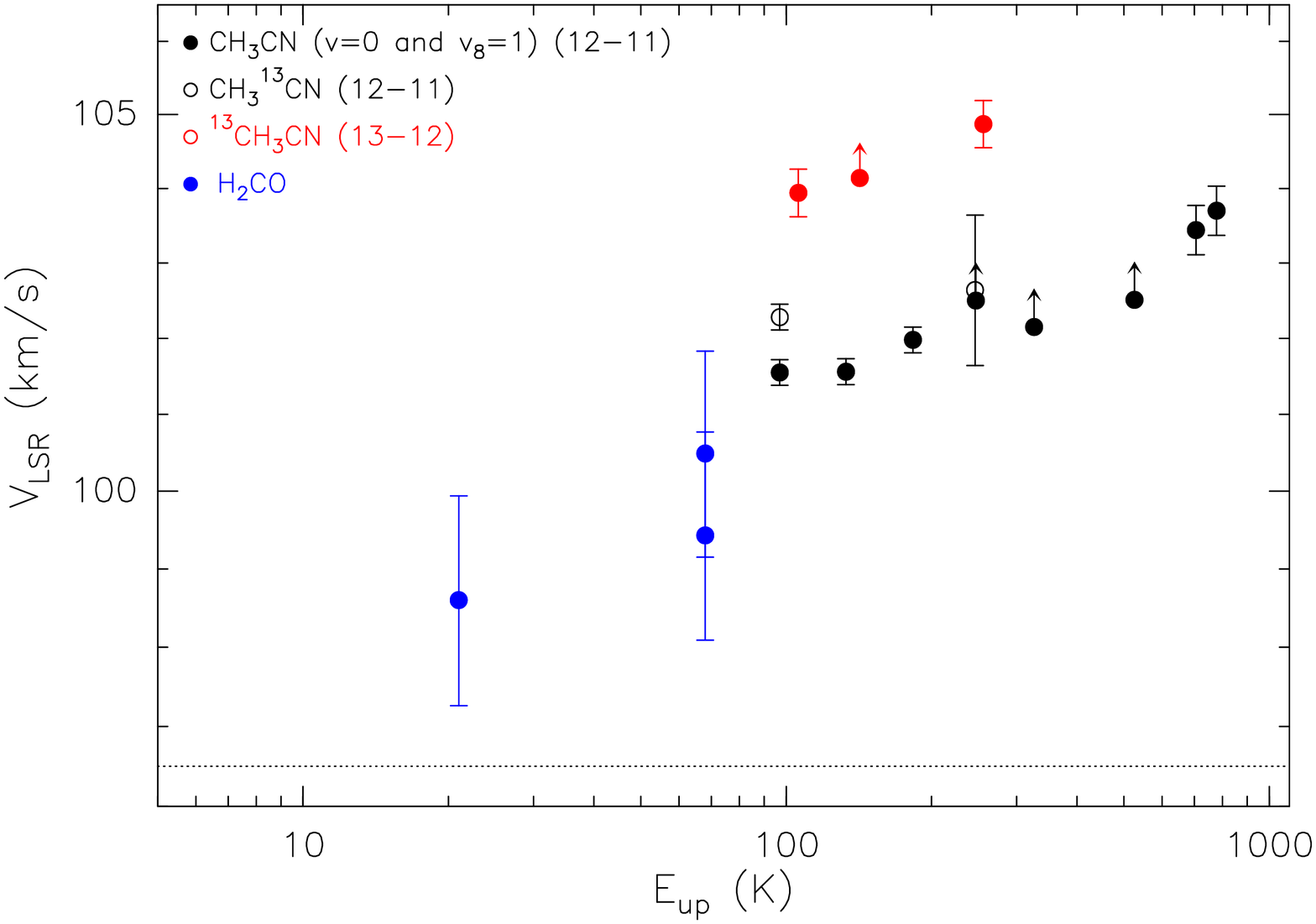}}
\caption{Velocity of the absorption feature (measured towards the dust continuum peak) versus the upper level energy of the corresponding \MCN, \MCNII, and \MCNI,   and H$_2$CO transitions. The systemic LSR velocity is 96.5 \kms\ and is indicated with a dotted line. The black arrows indicate lower limits to the velocity and are for those lines that are slightly blended with other lines at frequencies lower than their rest frequency.}
\label{fig-vred}
\end{figure}

\begin{table}
\caption[] {Upper level energy and velocity of the red-shifted absorption with respect to the systemic LSR velocity$^a$.}
\label{table-vred}
\begin{tabular}{lcc}
\hline
&\multicolumn{1}{c}{$E_{\rm up}$}
&\multicolumn{1}{c}{$|V_{\rm LSR}-V_{\rm red}|$}\\
\multicolumn{1}{c}{Line}
&\multicolumn{1}{c}{(K)} &
\multicolumn{1}{c}{(\kms)}
\\
\hline
H$_2$CO (3$_{0, 3} - 2_{0, 2}$) &21  &2.1$\pm$1.3 \\
H$_2$CO (3$_{2, 2} - 2_{2, 1}$)  &68 &4.0$\pm$1.3 \\
H$_2$CO (3$_{2, 1} - 2_{2, 0}$)  &68 &2.9$\pm$1.3\\
\MCN\    $K$=2  (12--11)  &97 &5.1$\pm$0.2  \\
\MCNII\    $K$=2  (12--11) &97 &5.8$\pm$0.2 \\
\MCNI\    $K$=2  (13--12) &106 &7.4$\pm$0.3  \\
\MCN\   $K$=3  (12--11) &133 &5.1$\pm$0.2  \\
\MCNI\   $K$=3  (13--12) &142 &$>7.6^b$  \\
\MCN\    $K$=4  (12--11) &183 &5.5$\pm$0.2  \\
\MCNII\    $K$=5  (12--11) &246 &6.1$\pm$0.2 \\
\MCN\    $K$=5  (12--11) &247 &$>6.0^b$  \\
\MCNI\    $K$=5  (13--12) &256 &8.4$\pm$0.3   \\
\MCN\    $K$=6  (12--11) &326 &$>5.7^b$  \\
\MCN\    $K$=8  (12--11) &526 & $>6.0^b$ \\
\MCN\ $K,l$=(3, $-1$) $v_8$=1 (12--11) &705 &6.9$\pm$0.3  \\
\MCN\  $K,l$=(6, 1) $v_8$=1 (12--11) &778 &7.2$\pm$0.3\\
\hline
\end{tabular}

 $^a$ The  velocity of the absorption feature has been measured towards the dust continuum peak and corresponds to that of the spectral channel with the lowest intensity at that position.   \\
  $^b$ Line slightly blended with other lines at frequencies lower than the rest frequency. \\
\end{table}

\begin{figure*}
\centerline{\includegraphics[angle=0,width=16cm,angle=0]{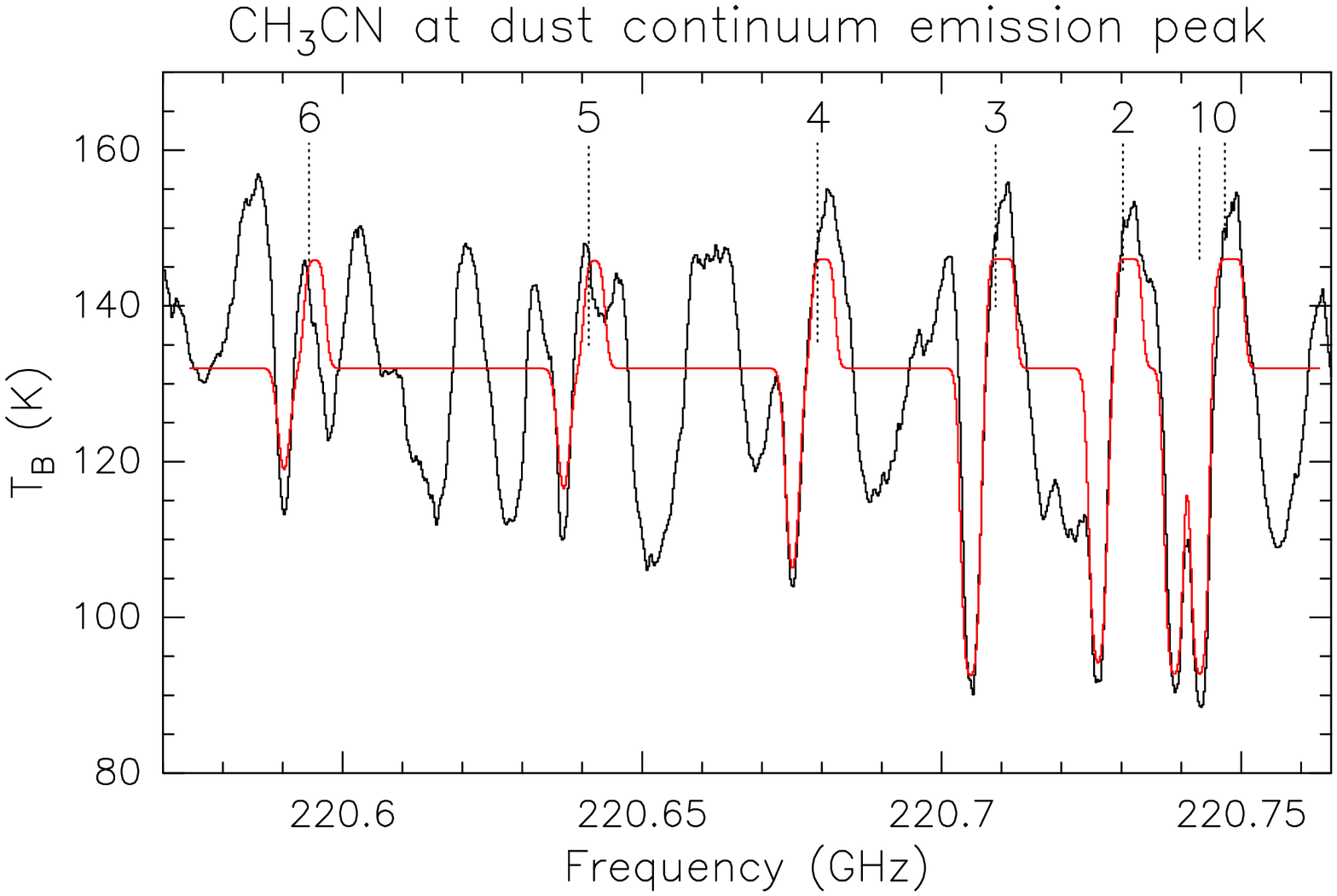}}
\caption{Spectrum obtained at the dust continuum emission peak of the G31 Main core ({\it black}) and best fit to the  \MCN\ (12--11) $K=0$ to 6 transitions ({\it red}). The spectrum has been modeled with different components: i) a background layer with homogeneous intensity to describe the dust continuum core, ii) a core layer to describe the emission, and iii) a foreground layer located in front of the core component to describe the absorption (see Sect.~\ref{excitation-absorption}). Note that the other emission and absorption lines not fitted correspond to other species.}
\label{fig-absorption-central}
\end{figure*}

\begin{figure*}
\centerline{\includegraphics[angle=0,width=16cm,angle=0]{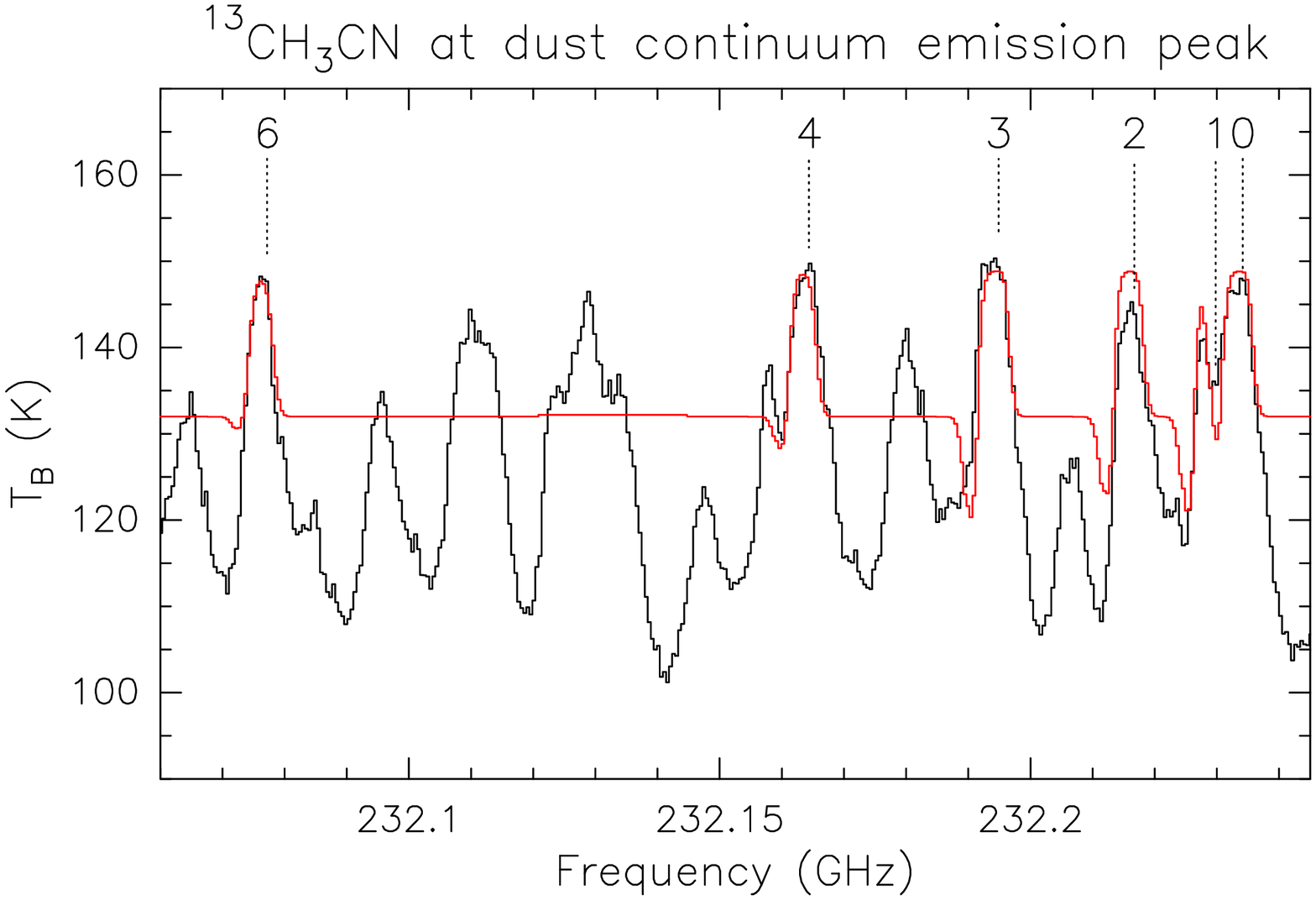}}
\caption{Same as Fig.~\ref{fig-absorption-central} for  \MCNI\ (12--11) $K=0$ to 6 transitions. }
\label{fig-absorption-central-13ch3cn}
\end{figure*}

To fit the line emission of \MCN\ and its isotopologues, the temperature of the continuum background layer was assumed to be constant and equal to the brightness temperature $T_{\rm B}=132$\,K measured at the dust continuum peak (see Table~\ref{table-cont}).  We assumed that the foreground or absorption component is extended and that it fills the beam (i.e., no beam dilution is considered).

We started by fitting \MCN\ and  \MCN\ $v_8=1$ (12--11), and  \MCNI\ (13--12) simultaneously. {\sc XCLASS} allows to simultaneously fit a species together with its isotopologues and vibrationally excited states by defining an isotopic ratio for each isotopologue. The transitions of \MCNII\ (12--11) were not modeled because they are too blended with those of  \MCN. Although we tried different approaches, like fixing the value of some parameters to reduce the number of free parameters, we could not find a satisfactory fit for all of them. This is probably due to the fact that the main species, the isotopologues, and the vibrationally excited states do not trace the same material and, thus, it is not possible to fit all of them with the same excitation conditions. Therefore, we decided to fit them separately. 

The first to be fitted was \MCN, because it is more clearly affected by absorption (in particular the lower energy transitions). As shown in Fig.~\ref{fig-spectra}, \MCN\ has been clearly detected up to the $K=8$ transition. For higher energy transitions ($K=9$ and 10), the blending with other species is so severe that it is not possible to estimate their contribution. We fitted the spectra up to the $K=6$ component (see Fig.~\ref{fig-absorption-central} and Table~\ref{table-fit}). The $K=7$ and 8 transitions could not be simultaneously fitted with the lower $K$ transitions, either because they are slightly blended with  some lines of \MCNII\ or because they could be tracing material with different excitation conditions than that traced by the lower energy transitions. 
To reduce the number of free parameters, we fixed the line width ($\Delta v$) and the offset velocity relative to the \Vlsr\ ($v_{\rm off}$) for both emission and absorption. Note that the fact that {\sc XCLASS} assumes that $v_{\rm off}$ is the same for all the transitions is questionable, because as we just discussed,  the velocity of the absorption feature increases with the energy of the transition (Table~\ref{table-vred} and Fig.~\ref{fig-vred}). As for $\Delta v$, it should also change with the energy of the transition, but in this case, the change is less evident and, therefore, the approximation of constant  $\Delta v$  is reasonable. We assumed that the source is extended and fills the beam size, 
like for the foreground component (the value obtained if the size parameter is left free is much larger than the beam size).  The best fit model indicates that the temperature of the emission layer (151\,K) is  higher than that of the dust emission (132\,K), while that of the absorption layer is lower (97\,K). As shown in Fig.~\ref{fig-absorption-central}, the \MCN\ modeled in emission is slightly saturated, as suggested by the fact that the $K=$0, 1, 2, and 3 transitions have almost the same brightness temperature despite their different energies. Therefore, the best fit column density of the emission component, $N_e=1\times10^{17}$\,cm$^{-2}$, should be taken as a lower limit. This  column density is one order of magnitude higher than that of the absorbing layer.

The isotopologue  \MCNI\ (13--12) has been clearly detected up to the $K=6$ transition. We fitted the spectra up to this transition avoiding the $K=5$ transition, which is highly blended with other species (see Fig.~\ref{fig-absorption-central-13ch3cn}). The source was assumed to be extended in both emission and absorption (i.e., no correction for beam dilution is applied). 
The fit to the absorption was highly dependent on the initial guesses and  it is not as good as that to the absorption in \MCN\ (see Figs.~\ref{fig-absorption-central} and \ref{fig-absorption-central-13ch3cn}). Therefore, in Table~\ref{table-fit-13ch3cn} we only give the best fit parameters of the emission, which have been obtained by fixing the line width and the velocity offset.  
The excitation temperature of the emission is similar, within the uncertainties, to that obtained for \MCN. Regarding the column density of \MCNI, the value obtained is only $\sim$2 times lower than  that of \MCN.
This indicates a very low [\MCN]/[\MCNI] relative abundance, especially if one takes into account that the relative abundance of [$^{12}$C/$^{13}$C] at the galactocentric distance of 4.5 kpc of G31 would be 47 (Milam et al.~\cite{milam05}). Note that the column density of  \MCN\ in emission should be taken as a lower limit because the emission is saturated. This could explain in part the low values obtained in emission for the [$^{12}$C/$^{13}$C] ratio, although the \MCNI\  emission also seems slightly saturated (see Fig.~\ref{fig-absorption-central-13ch3cn}).

We cannot disregard the possibility that the [$^{12}$C/$^{13}$C] ratio in the G31 core is unusually low. To investigate this possibility we fitted both the \MCN\ and \MCNI\ emission in a ring located at a distance of $1\farcs32$ from the center of the core, where the absorption is negligible.  For this ring, the temperature of the background layer was assumed to be 1.8\,K, which is the  continuum brightness  temperature at that radius.  The best fit indicates a similar excitation temperature, within the errors, for both \MCN\ and \MCNI, but a column density a factor 10 lower ($\log N_e = 14.3^{-0.2}_{+1.7}$ cm$^{-2}$) for  \MCNI\ than for \MCN\ ($\log N_e = 15.3^{-0.01}_{+1.3}$ cm$^{-2}$). A  [$^{12}$C/$^{13}$C]  ratio of 10 is greater than the value obtained towards the central position but still significantly less than the typical value. To check wether this low ratio is also seen in other species, we estimated the column densities of \MF\ and CH$_3$O$^{13}$CHO in the rings not affected by absorption and obtained [$^{12}$C/$^{13}$C]  $\simeq$10. This indicates that in the G31 core there is an overabundance of $^{13}$C with respect to $^{12}$C.

The  \MCN\ $v_8=1$ (12--11) transitions at the central position have been fitted only with an emission component  because an additional foreground absorbing layer 
produces a negligible change in the fit. This is consistent with the fact that the red-shifted absorption is less evident for these high energy transitions. The best fit parameters are given in Table~\ref{table-fit-v8}.  The excitation temperature of the vibrationally excited transitions is consistent with that obtained for \MCN\ and \MCNI, but the column density is almost one order of magnitude higher than that of the ground state.  Therefore, we decided to fit the  \MCN\ $v_8=1$ emission by fixing 
$N_e$ to a value of $1\times10^{17}$\,cm$^{-2}$, obtained for the ground state \MCN, and got $T_e =  224$\,K. 
Despite the large uncertainties in the fit, the fact that one needs either a much higher column density or a much higher excitation temperature to model the vibrationally excited transitions suggests that these high energy transitions are likely tracing hotter and denser material located deeper in the core.  

Because of the existence of temperature and density gradients in the G31 core, as discussed in  Sect.~\ref{t-gradient}, the \MCN\ $v_8=1$ emission cannot be well represented by the same excitation conditions as those of the ground state emission. Radiative pumping by IR photons may also
play a role in the excitation of \MCN\ $v_8=1$ (e.g., Goldsmith et al.~\cite{goldsmith83}). To carry out an analysis of collisional excitation and check whether collisional processes could produce the observed excitation of the $v_8$ transitions it is necessary to know the spontaneous decay rate,  $A_{\rm vib}$, and the collision cross sections for vibrational excitation $<\sigma v>_{\rm vib}$. Using the estimates of $A_{\rm vib}$$= 1.6\times10^{-2}$\,s$^{-1}$ of Koivusaari et al.~(\cite{koivusaari92}), and assuming the same collisional excitation rate as for the vibrational ground state,  $<\sigma v>_{\rm vib}$=$2\times10^{-10}$\,cm$^3$\,s$^{-1}$ (Green \cite{green86}), then collisional excitation is possible only for $n\gtrsim10^{8}$\,cm$^{-3}$. Because $n$ estimated in the core ranges from $5.4\times10^6$ cm$^{-3}$ for $R=1\farcs076$ to $\simeq1\times10^8$ cm$^{-3}$ for $R=0\farcs22$, we conclude that the vibrational levels could be radiatively pumped.  The first vibrationally excited state $v_8=1$ is at 525\,K above the \MCN\ ground vibrational state and can be pumped by 27\,$\mu$m photons.

\begin{table}
\caption[] {Best fit physical parameters for the \MCN\ (12--11) $K=0$ to 6  transitions towards the center of the G31 core.}
\label{table-fit}
\begin{tabular}{cc}
\hline
\multicolumn{1}{c}{Emission} 
&\multicolumn{1}{c}{Absorption} \\
\hline
$T_e$ =  $151^{-8}_{+83}$  K & $T_a$ = $97^{-4}_{+3}$ K \\ 
$\log N_e > 17$ cm$^{-2}$ & $\log N_a = 15.9^{-0.2}_{+0.3}$ cm$^{-2}$ \\
$\Delta v_e^{\, \rm a} = 3$ \kms\ & $\Delta v_a ^{\, \rm a} = 3$ \kms\ \\
$v_{{\rm off}\,e}^{\, \rm a} = -1.3$ \kms\ & $v_{{\rm off}\,a}^{\, \rm a} = 5.5$ \kms\  \\
\hline
\end{tabular}
\\
$^{\rm a}$ Parameter fixed.
\end{table}

\begin{table}
\caption[] {Best fit  physical parameters for the \MCNI\ (12--11) $K=0$ to 6  transitions towards the center of the G31 core.}
\label{table-fit-13ch3cn}
\begin{tabular}{cc}
\hline
\multicolumn{1}{c}{Emission} 
\\
\hline
$T_e$ =  $155^{-50}_{+107}$  K \\
$\log N_e = 16.8^{-1.2}_{+0.1}$ cm$^{-2}$ \\
$\Delta v_e^{\, \rm a} = 3.5$ \kms\ \\
$v_{{\rm off}\,e}^{\, \rm a} = 1.3$ \kms\ \\
\hline
\end{tabular}
\\
$^{\rm a}$ Parameter fixed.
\end{table}

\begin{table}
\caption[] {Best fit  physical parameters for the \MCN\ $v_8=1$ (12--11) transitions towards the center of the G31 core.}
\label{table-fit-v8}
\begin{tabular}{cc}
\hline
\multicolumn{1}{c}{Emission} 
&\multicolumn{1}{c}{Emission}
\\
& (fixing $N_e$)
\\
\hline
$T_e$ =  $165^{-153}_{+64}$  K & $T_e$ = $224^{-208}_{+3}$ K \\ 
$\log N_e = 17.7^{-2.9}_{+1.24}$ cm$^{-2}$ & $\log N_e^{\, \rm a} = 17$ cm$^{-2}$ \\
$\Delta v_e^{\, \rm a} = 6$ \kms\ & $\Delta v_e ^{\, \rm a} = 6$ \kms\ \\
$v_{{\rm off}\,e} = 0.5^{-2.7}_{+5.2}$ \kms\ & $v_{{\rm off}\,e} = 0.6^{-2.3}_{+0.15}$ \kms\  \\
\hline
\end{tabular}
\\
$^{\rm a}$ Parameter fixed.
\end{table}

\subsection{A monolithic core?}

The most striking characteristic of the \g31 Main core is its compact appearance. As seen in Fig.~\ref{fig-cont}, the dust continuum emission does not show any hint of fragmentation although the core is well resolved in our maps.  
The presence of red-shifted absorption in the core would suggest that what we are observing in \g31 is a ``real'' massive core undergoing monolithic collapse, as suggested by McKee \& Tan~(\cite{mckee02}).   However, one cannot discard the possibility that this homogeneous and monolithic appearance is due to a combination of large opacity of the 217\,GHz continuum emission and insufficient angular resolution. One may argue that the size of a typical fragment should be of the same order as the separation between the two free-free sources detected inside the core, which is  $\sim$$0\farcs2$ or $\sim$1600\,au. By using Eqs.~(\ref{tex}) and (\ref{rho}), we have estimated the temperature and volume density in the inner part of the core ($R\leq0\farcs22$), and the thermal Jeans length, and found that it ranges from $\sim$1300 to 2000\,au. Therefore, it is comparable to the separation of $\sim$1600\,au of the free-free sources.
This implies that fragmentation, if present as expected from theoretical calculations (see Peters et al.~\cite{peters10b}), cannot be properly imaged with the current observations.   

One  must also take into account that the continuum emission observed by us at 217\,GHz is optically thick (the brightness temperature of the continuum peak, $\sim$132 K, is comparable to the dust temperature, $\gtrsim$150 K). The high opacity of G31 has already been determined  by Rivilla et al.~(\cite{rivilla17}) based on multi-wavelength observations. Such a large opacity ($\lesssim$2) is likely to mask any inhomogeneity of the core. This result is similar to what is found in the most massive core of SgrB2(N). Sub-arcsecond resolution observations with the SMA (345 GHz; Qin et al.~\cite{qin11}) and with ALMA (250 GHz; S\'anchez-Monge et al.~\cite{sanchez-monge17}) revealed a core with monolithic appearance and a radius of about 6000~au. In this case, 3D radiative transfer modelling of the region (Schmiedeke et al.~\cite{schmiedeke16}) suggests that the emission at frequencies above 200 GHz is optically thick when observed at an angular resolution $<$$0.5''$.

These results altogether pose questions about the nature of the G31 Main core. On the one hand, Girart et al.~(\cite{girart09}) have revealed the presence of a magnetic field uniformly threading the core, with a strength in the plane of the sky of $\sim$10\,mG. This, together with the linear increase of rotation velocity with radius (suggested by each PV plot in Fig.~\ref{fig-pv}), and the (apparent) homogeneity of the core, point to a monolithic core stabilized by the magnetic field and undergoing solid-body rotation. On the other hand, a mass-to-magnetic flux ratio of $\sim$2.7 times the critical value for collapse (Girart et al.~\cite{girart09}), the presence of red-shifted absorption, 
the existence of two embedded massive stars (the two free-free sources),  
and the rotational spin-up towards the core center (suggested by Fig.~\ref{fig-pv-div}) are consistent with an unstable core undergoing fragmentation with infall and differential rotation due to conservation of angular momentum. We believe that an optically thick core with unresolved clumpiness and undergoing infall and rotation spin-up can explain all these features, as previously discussed. 

Future approved ALMA observations at a much higher angular resolution of $\sim$$0\farcs08$, corresponding to $\sim$600\,au at the distance of the source, will help us to reveal the existence of clumpiness and thus confirm whether G31 is indeed a dynamically collapsing core undergoing global fragmentation.

\section{Conclusions}

As part of our continuing effort to search for circumstellar disks around high-mass young stellar objects, we have observed the O-type star-forming region G31.41+0.31 with ALMA at 1.4\,mm. The angular resolution ($\sim$$0\farcs2$) of the observations has allowed us to study the G31 HMC in unprecedented detail, potentially tracing structures as small as $\sim$1600\,au.  

The dust continuum emission  has been resolved into two cores, a Main core chemically rich that peaks close to the position of the two unresolved free-free continuum sources detected by Cesaroni et al.~(\cite{cesa10}), and a much weaker and smaller NE core. The dust continuum emission of the Main core looks featureless in our observations, although it has an average half-power diameter of $\sim$5300\,au, much greater than our resolution ($\sim$1600\,au). 

The SiO emission reveals the presence of at least three outflows associated with the G31 core: an E--W outflow to the south of the dust continuum emission peak of the Main core and not associated with any of the two embedded free-free sources; a N--S outflow associated with water masers, which could be driven by the millimeter source associated with the dust continuum emission peak of the Main core or by one of the two free-free sources embedded in this core; and a NE--SW outflow to the south of the Main core and not associated with it.

The high sensitivity of the ALMA observations has allowed us  to detect several transitions of the high-density tracer \MF, both ground state and torsionally excited, and this has allowed us to estimate the physical parameters of the Main core. The LTE modeling of the \MF\ emission indicates that there is an excitation temperature  gradient in the core, with $T_{\rm ex}\propto R^{-0.77}$, consistent with the profiles of hot molecular cores modeled by Nomura \& Millar~(\cite{nomura04}) and Osorio et al.~(\cite{osorio09}). The \MF\ column density ($N_{\rm CH_3OCHO}$) profile in the core is consistent with a broken  power law, with the inner part of the core having $N_{\rm CH_3OCHO}\propto R^{-0.82}$, and the outer part being considerably  steeper, with   $N_{\rm CH_3OCHO}\propto R^{-4.63}$. On the other hand, the H$_2$ column density profile, estimated from the dust continuum emission, does not show any sharp drop in the outer part of the core, with $N_{\rm H_2}\propto R^{-1.15}$. We conclude that the drop in  \MF\ abundance is real. This suggests that \MF\ is concentrated in the inner regions ($R<5000$\,au) of the G31 Main core, and confirms that complex organic molecules are reliable tracers of the gas located close to the forming protostars. The mass of the G31 Main core, estimated assuming that the dust emission is in the Rayleigh-Jeans
regime and that the density and temperature of the core follow the previously established power-law distributions, is $\sim$120\,$M_\odot$, while the mean volume density at the radius of the core (8500\,au) is $\sim$5.4$\times10^6$\,cm$^{-3}$.

The emission of \MCN\ and \MF, both ground state and vibrationally excited, and of \MCN\ isotopologues in the Main core shows a clear velocity gradient along the NE--SW direction consistent with what was previously found by Beltr\'an et al.~(\cite{beltran04}) and Cesaroni et al.~(\cite{cesa11}).  The velocity appears to increase linearly with distance from the core center, consistent with solid-body rotation. However, the PV plots along the major axis of the core indicate that the velocities increase for increasing energies. This suggests that the rotation spins up towards the center, as expected for conservation of angular momentum in a rotating and infalling structure, which appears initially inconsistent with the solid-body rotation.  A possible explanation for this could be that each line is tracing only a relatively narrow annulus of material. The projection of the rotation velocity along the line of sight increases linearly with distance from the star, mimicking solid-body rotation, while at the same time, higher energy lines trace smaller annuli, closer to the star and thus rotating faster.

The line emission towards the dust continuum peak of the Main core shows inverse P-Cygni profiles in basically all the species and at all energies, although  the H$_2$CO and lowest-energy \MCN\ transitions show the strongest absorption. The detection of red-shifted absorption towards the core center supports the existence of infall in the core. The velocity of the absorption feature (which is a good proxy for the infall velocity) becomes progressively more red shifted with increasing energy of the transition. This suggests that the infall is accelerating towards the center of the core, consistent with free-fall collapse. The excitation conditions of the emitting and absorbing  gas components at the position of the dust continuum emission peak have been estimated by modeling the \MCN\ and \MCNI\ transitions, assuming LTE. We find that the column density of the absorbing layer is  more than one order of magnitude smaller than that of the emitting layer. We also find that the [$^{12}$C/$^{13}$C] ratio in the G31 core is unusually low, suggesting that there is an overabundance of $^{13}$C with respect to $^{12}$C. The modeling of the \MCN\ $v_8=1$ emission at the central position indicates that to excite these transitions either a much higher column density or a much higher excitation temperature than that needed to populate the ground state levels of \MCN\ is necessary. This suggests that \MCN\ $v_8=1$  could be excited via radiative pumping by mid-IR (27$\mu$m) photons. 

Although its homogeneous appearance suggests that G31 Main could be a true monolithic core, as predicted by  McKee \& Tan~(\cite{mckee02}), the presence of red-shifted absorption, the existence of two embedded massive stars at the center, and the rotational spin-up are consistent with an unstable core undergoing fragmentation with infall and differential rotation due to conservation of angular momentum. Therefore, we argue that the most likely explanation for its morphology is that the dust emission is optically thick, as suggested by the detection of red-shifted absorption, and this prevents the detection of any inhomogeneity in the core.

\begin{acknowledgements}  

 We thank the referee, Dr.\ Paul Goldsmith, for the careful reading of the manuscript.
This paper makes use of the following ALMA data: ADS/JAO.ALMA\#2013.1.00489.S. ALMA is a partnership of ESO (representing
its member states), NSF (USA), and NINS (Japan), together with NRC (Canada), and NSC and ASIAA (Taiwan), in cooperation with the Republic
of Chile. The Joint ALMA Observatory is operated by ESO, AUI/NRAO, and
NAOJ.  This work was partly supported by the Italian Ministero dell'Istruzione, Universit\`a e Ricerca through the grant Progetti Premiali 2012 -- iALMA (CUP C52I13000140001), by the Deutsche Forschungs-Gemeinschaft (DFG, German Research Foundation) - Ref no. FOR 2634/1 TE 1024/1-1, and by the DFG cluster of excellence Origin and Structure of the Universe (\url{http://www.universe-cluster.de}).
V.M.R. has received funding from the European Union's Horizon 2020 research and innovation programme under the Marie Sk\l{}odowska-Curie grant agreement No 664931.
\'A.S.M. and P.S. are partially supported by the Deutsche Forschungsgemeinschaft through grant SFB 956 (subproject A6).  XCLASS development is supported by  BMBF/Verbundforschung through the Projects ALMA-ARC 05A11PK3 and 05A14PK1 and through  ESO Project 56787/14/60579/HNE. 
A.A., H.B., and J.C.M. acknowledge support from the European Research Council under the European Community's Horizon 2020 framework program (2014-2020) via the ERC Consolidator grant ``From Cloud to Star Formation (CSF)'' (project number 648505). 
R.G.M. acknowledges support from UNAM--PAPIIT program IA102817.
A.K. and R.K. acknowledge financial support via the Emmy Noether Research Programme funded by the German Research Foundation (DFG) under grant no. KU 2849/3-1.
M.S.N.K. acknowledges the support from Funda\c{c}\~ao para a Ci\^encia e Tecnologia (FCT) through Investigador FCT contracts IF/00956/2015/CP1273/CT0002, and the H2020 Marie-Curie Intra-European Fellowship project GESTATE (661249).

 \end{acknowledgements}

\end{document}